\documentclass[fleqn,usenatbib]{mnras}

\usepackage{newtxtext,newtxmath}

\usepackage[T1]{fontenc}

\DeclareRobustCommand{\VAN}[3]{#2}
\let\VANthebibliography\thebibliography
\def\thebibliography{\DeclareRobustCommand{\VAN}[3]{##3}\VANthebibliography}

\usepackage{graphicx}	
\usepackage{amsmath}	
\usepackage{xcolor}
\usepackage{pdflscape}
\usepackage{orcidlink}
\usepackage{tablefootnote}

\newcommand*\Phosphoros{\texttt{Phosphoros} \,}
\newcommand*\EAZY{\texttt{EAzY} \,}
\newcommand*\photutils{\texttt{photutils} \,}

\title[Spectroscopic redshifts in SMACS~J0723.3$-$7327]{The first large catalogue of spectroscopic redshifts in Webb's First Deep Field, SMACS~J0723.3$-$7327}
\author[Noirot et al.]
{Ga\"el Noirot$^{1}$\thanks{Contact e-mail: \href{mailto:Gael.Noirot@smu.ca}{Gael.Noirot@smu.ca}},
Guillaume Desprez$^{1}$\orcidlink{0000-0001-8325-1742},
Yoshihisa Asada$^{1,2}$\orcidlink{0000-0003-3983-5438},
Marcin Sawicki$^{1}$\orcidlink{0000-0002-7712-7857},
Vicente Estrada-Carpenter$^{1}$\orcidlink{0000-0001-8489-2349},\newauthor
Nicholas Martis$^{1,3}$\orcidlink{0000-0003-3243-9969},
Ghassan Sarrouh$^{4}$\orcidlink{0000-0001-8830-2166},
Victoria Strait$^{5}$\orcidlink{0000-0002-6338-7295},
Roberto Abraham$^{6,7}$\orcidlink{0000-0002-4542-921X},
Maruša Bradač$^{8}$\orcidlink{0000-0001-5984-0395},\newauthor
Gabriel Brammer$^{5}$\orcidlink{0000-0003-2680-005X},
Kartheik Iyer$^{6,9}$\orcidlink{0000-0001-9298-3523},
Shannon MacFarland$^{1}$,
Jasleen Matharu$^{5}$\orcidlink{0000-0002-7547-3385},
Lamiya Mowla$^{6}$\orcidlink{0000-0002-8530-9765},\newauthor
Adam Muzzin$^{4}$,
Camilla Pacifici$^{10}$\orcidlink{0000-0003-4196-0617},
Swara Ravindranath$^{10}$\orcidlink{0000-0002-5269-6527},
Chris J.~Willott$^{3}$\orcidlink{0000-0002-4201-7367},
Lo\"ic Albert$^{11}$\orcidlink{0000-0003-0475-9375},\newauthor
Ren\'e Doyon$^{11}$\orcidlink{0000-0001-5485-4675},
John B.~Hutchings$^{3}$,
and Neil Rowlands$^{12}$\orcidlink{0000-0002-1715-7069}\\
$^{1}$Institute for Computational Astrophysics and Department of Astronomy \& Physics, Saint Mary's University, 923 Robie Street, Halifax, NS B3H 3C3, Canada\\
$^{2}$Department of Astronomy, Kyoto University, Sakyo-ku, Kyoto 606-8502, Japan\\
$^{3}$National Research Council of Canada, Herzberg Astronomy \& Astrophysics Research Centre, 5071 West Saanich Road, Victoria, BC, V9E 2E7, Canada\\
$^{4}$Department of Physics and Astronomy, York University, 4700 Keele St. Toronto, Ontario, M3J 1P3, Canada\\
$^{5}$Cosmic Dawn Center (DAWN), Niels Bohr Institute, University of Copenhagen, Jagtvej 128, DK-2200 Copenhagen N, Denmark\\
$^{6}$Dunlap Institute for Astronomy and Astrophysics, 50 St.~George Street, Toronto, Ontario, M5S 3H4, Canada\\
$^{7}$David A.~Dunlap Department of Astronomy and Astrophysics, University of Toronto, 50 St.~George Steet, Toronto, Ontario, M5S 3H4, Canada\\
$^{8}$Department of Mathematics and Physics, Jadranska ulica 19, SI-1000 Ljubljana, Slovenia\\
$^{9}$Columbia Astrophysics Laboratory, Columbia University, 550 West 120th Street, New York, NY 10027, USA\\
$^{10}$Space Telescope Science Institute, 3700 San Martin Drive, Baltimore, Maryland 21218, USA\\
$^{11}$Institut Trottier de Recherche sur les Exoplan\`etes (iREx), Universit\'e de Montr\'eal, D\'epartement de Physique, C.P. 6128 Succ. Centre-ville, Montr\'eal, QC H3C 3J7,\\ Canada\\
$^{12}$Honeywell Aerospace, 303 Terry Fox Drive, Ottawa, ON K2K 3J1, Canada\\
}

\date{Accepted XXX. Received YYY; in original form ZZZ}

\pubyear{2022}

\begin{document}
\label{firstpage}
\maketitle
\pagerange{\pageref{firstpage}--\pageref{lastpage}}

\begin{abstract}
%
We present a spectroscopic redshift catalogue of the  SMACS~J0723.3$-$7327 field (``Webb's First Deep Field'') obtained from {\it JWST}/NIRISS grism spectroscopy and supplemented with {\it JWST}/NIRSpec and VLT/MUSE redshifts. The catalogue contains a total of 190 sources with secure spectroscopic redshifts, including 156 NIRISS grism redshifts, 123 of which are for sources whose redshifts were previously unknown. These new grism redshifts are secured with two or more spectroscopic features (64 sources), or with a single spectral feature whose identity is secured from the object's nine-band photometric redshift (59 sources). These are complemented with 17 NIRSpec and 48 MUSE redshifts, including six new NIRSpec redshifts identified in this work. In addition to the 
$z_{\rm cl}=0.39$ cluster galaxy redshifts (for which we provide $\sim$40 new NIRISS absorption-line redshifts), we also find three prominent galaxy overdensities at higher redshifts -- at $z=1.1$, $z=1.4$, and $z=2.0$ -- that were until now not seen in the {\it JWST}/NIRSpec and VLT/MUSE data. The paper describes the characteristics of our spectroscopic redshift sample and the methodology we have employed to obtain it. Our redshift catalogue is made available to the community at {\url{https://niriss.github.io/smacs0723}}. 
\end{abstract}




\section{Introduction} \label{sec:intro}

Understanding how galaxies form and evolve over cosmic time is at the core of two of the four strategic goals that motivated the construction of {\it JWST}, launched on 2021 December 25 after many years of development. 
The first image released to the scientific community is that of the gravitational-lensing galaxy cluster SMACS~J0723.3$-$7327 \citep[``Webb's First Deep Field'',][]{Pontoppidan2022}, made public on 2022 July 11. These first data demonstrate the unprecedented power of {\it JWST} for near- and mid-infrared extragalactic astronomy, and have already formed the basis of a number of early science papers \citep[e.g.,][to give a few examples]{Atek2022, Caminha2022, Ferreira2022, Mowla2022, Schaerer2022, Yan2022, Carnall2022, Curti2023}.
The SMACS~J0723.3$-$7327 field is thus rapidly becoming a major public resource for research in the field of galaxy formation and evolution.

Accurate galaxy redshifts are vital for galaxy evolution studies as well as for the modeling of the gravitational-lensing potential of the cluster. While a number of redshifts in the SMACS~J0723.3$-$7327 field has already been published from ground-based studies \citep[VLT/MUSE, ][]{Caminha2022}, and a few from the analysis of the public {\it JWST}/NIRSpec observations of the field \citep[e.g., ][]{Schaerer2022, Carnall2022}, a much more comprehensive redshift catalogue is both desirable and possible with the existing {\it JWST} dataset, particularly using the {\it JWST}/NIRISS slitless grism spectroscopy data of the Early Release Observations (EROs) of the field \citep{Pontoppidan2022}.

We here develop such a catalogue using primarily the {\it JWST}/NIRISS slitless grism spectroscopy of the field, supported with {\it JWST}/NIRCam and {\it HST}/ACS photometry. The NIRISS grism dataset in SMACS~J0723.3$-$7327 was observed with two {broad-band filters (F115W and F200W). In particular, it does not include the F150W filter between the F115W and F200W}, limiting the yield of secure redshifts with multiple spectral features. Nevertheless, we are able to securely identify $156$ NIRISS grism redshifts in the field, $123$ of which are for sources whose redshifts were previously unknown.  Among these $156$ (123) secure NIRISS grism redshifts, $87$ (64) are identified unambiguously through the identification of multiple spectral features, and a further $69$ (59) through a single spectral feature whose nature is unambiguously determined based on the object's nine-band photometric redshift.
While the NIRISS EROs of the SMACS~J0723.3$-$7327 field do not fully demonstrate the power of NIRISS for obtaining large redshift samples due to the lack of the F150W filter in these observations, this should be alleviated in NIRISS programs that include the F150W channel in addition to the F115W and the F200W, such as the Canadian NIRISS Unbiased Cluster Survey (CANUCS) project\footnote{Information on the CANUCS Guaranteed Time Observations program can be found at {\url{https://niriss.github.io}.}}. 

In addition to the NIRISS slitless grism data, we also re-analyse the public {\it JWST}/NIRSpec multi-object spectroscopic slit data and extract six additional, previously unpublished, secure redshifts that we include in our catalogue.  
Throughout this work, we take a conservative approach to obtain secure redshifts for a fairly large subset of sources rather than a complete catalogue for the full sample. Objects with redshifts that we consider to be uncertain are thus excluded from our final catalogue. 
We make our spectroscopic redshift catalogue public\footnote{The catalogue is made available at {\url{https://niriss.github.io/smacs0723}}.}, and the purpose of this paper is to describe the techniques we use to create it and give some characteristics of the data it contains.

This paper is structured as follows. Section~\ref{sec:data} describes our data and the basic processing we use to reduce it, including the basic reduction of the {\it JWST}/NIRISS data. In Section~\ref{sec:methods}, we describe our methods to ({\it i}) derive initial NIRISS grism redshifts (Sec.~\ref{sec:methods-grismredshifts}), ({\it ii}) obtain our photometric catalogue (Sec.~\ref{sec:photometry}) and photometric redshifts of the sources (Sec.~\ref{sec:photo_z}), and ({\it iii}) obtain our final spectroscopic redshift catalogue of the field (Sec.~\ref{sec:methods-grism-photoz}). Section~\ref{sec:redshift_sample} presents the properties of our redshift sample, including the fidelity of our NIRISS grism redshifts (Sec.~\ref{sec:redshift_fidelity}), the redshift distribution of the sample (Sec.~\ref{subsec:redshift-distribution}), and the colour distribution of our sources (Sec.~\ref{sec:colors}). In Section~\ref{sec:caveats} we present the limitations and caveats of this work before summarizing our work and presenting our conclusions in Section \ref{sec:conclusion}. 

Throughout, we use AB magnitudes, and a flat $\Lambda$CDM cosmology with $\Omega_{\rm M}=0.3$, $\Omega_{\rm \Lambda}=0.7$, and $H_0=70~{\rm km\,s^{-1}\,Mpc^{-1}}$.

\section{Data and basic processing} \label{sec:data}

We make use of the {\it JWST}/NIRCam images, NIRISS grism spectra, and NIRSpec multi-slit spectra released as part of the {\it JWST} Early Release Observations (EROs; \citealt{Pontoppidan2022}) of the field of the $z_{\rm cl}=0.39$ cluster SMACS~J0723.3$-$7327. We also use archival {\it HST} Advanced Camera for Surveys (ACS) imaging of the field from the RELICS program \citep{Coe2019}.
We additionally supplement our data with published VLT/MUSE spectroscopic redshifts \citep{Caminha2022}, where available. We describe these data and their basic processing in the following sections.

\subsection{{\it JWST}/NIRCam and {\it HST}/ACS imaging data}\label{sec:nircam_hst_imaging}

The {\it JWST}/NIRCam images were obtained through the six broad-band filters F090W, F150W, F200W, F277W, F356W, and F444W, while the {\it HST}/ACS imaging was obtained in the F435W, F606W and F814W broad-band filters, respectively.
The exposure time of the NIRCam data is $7.5$~ks, potentially reaching point-source depths of $\rm AB\sim30$~mag over much of the field \citep{Pontoppidan2022}.
The observations were taken using the {\tt INTRAMODULEX} dither pattern with nine dither positions and the {\tt MEDIUM8} readout pattern \citep{Pontoppidan2022}.

We first extract the {\it JWST}/NIRCam uncalibrated ramp exposures of the EROs of the field from the Mikulski Archive for Space Telescopes (MAST\footnote{\url{https://archive.stsci.edu/}}), and first run a modified version of the {\tt Detector1Pipeline} ({\tt calwebb\_detector1}) stage of the official STScI pipeline. 
This initial step includes detector-level correction and pixel flagging (including flagging of the ``snowball'' artifacts), as well as an additional correction of the $1/f$-noise. The ramp fitting, dark current subtraction, persistence correction, saturation flagging and cosmic ray rejection are also included in this step, which produces the ``rate images''. 
The {\it JWST} Operational Pipeline
({\tt CRDS\_CTX}) we use to reduce the NIRCam data
is {\tt jwst\_0916.pmap}. This is a pre-flight reference file for which photometric zero-points may be incorrect \citep[see, e.g.,][]{Boyer2022}. We therefore use {\tt EAzY} \citep{Brammer2008} to rederive proper photometric zero-points consistent with the photometric redshift fitting of our source catalogue as we describe in \citet{Mowla2022}. {Note that, as a consistency check, we use {\tt Phosophoros} (\citealt{Desprez2020}; Paltani et al., in prep) to re-derive photometric zero-points {\it a posteriori} based on the new spectroscopic redshifts derived in our work. These updated zero-points do not change our results (see Sec.~\ref{sec:caveats} for details) and we therefore only use the {\tt EAzY} photometric zero-points in our analysis.}

After this initial step, we use version 1.6.0 of the grism redshift and line analysis software for space-based spectroscopy \citep[{\tt Grizli};][]{Brammer2021} to perform the remaining steps of the data reduction. Specifically, we first perform the astrometric alignement of the different exposures to the {\it HST} images, then subtract the sky-background from the exposures, and drizzle the {\it JWST}/NIRCam and {\it HST}/ACS imaging to a common pixel scale of $0\farcs04$. Additionally to these steps, we model and subtract the intracluster light and the light from the brightest cluster galaxies using a custom code (Martis et al., in preparation), and we PSF-match our imaging to the lowest resolution image (F444W) using empirical PSFs constructed from bright stars in the field.

\subsection{NIRISS Wide-Field Slitless Spectroscopy}

\subsubsection{Data and basic processing}\label{sec:grism_reduction}

Additionally to the imaging data, the SMACS~J0723.3$-$7327 field was observed with the Wide-Field Slitless Spectroscopy (WFSS) mode \citep{willott2022} of the Near-InfraRed Imager and Slitless Spectrograph \citep[NIRISS,][Doyon et al.~in prep]{Doyon2012}. The NIRISS WFSS observations used the two orthogonal, low-resolution ($R\sim 150$), grisms, GR150C and GR150R, and were taken through the F115W ($\lambda=1.0-1.3$~\micron) and F200W ($\lambda=1.75-2.25$~\micron) filters. Total integrations were $2.8$~ks per channel, with an eight-point medium dither pattern \citep{Pontoppidan2022}.

The basic processing steps of the NIRISS exposures follow the same procedure as done for the NIRCam imaging data. In short, we extract the uncalibrated data from MAST, run a modified version of the official {\tt Detector1Pipeline} ({\tt calwebb\_detector1}) for detector-level flagging and corrections including the $1/f$-noise, and then use {\tt Grizli} to perform the additional reduction steps (namely, exposure alignment to the {\it HST} images, sky-background subtraction, and drizzling of the exposures). The {\it JWST} Operational Pipeline
({\tt CRDS\_CTX}) we use to reduce the NIRISS data
is {\tt jwst\_0932.pmap}.
The {\tt Grizli} reduction steps of the NIRISS data also include the creation of a NIRISS direct image mosaic from which diffraction spikes of bright sources are masked.
Source detection is then performed on the NIRISS mosaic image with the Source Extractor \citep{Bertin1996} python wrapper {\tt sep} \citep{Barbary2016}, using the default detection parameters implemented in {\tt Grizli} (a detection threshold `{\tt threshold}' of $1.8\sigma$ above the global background RMS, a minimum source area in pixels `{\tt minarea}' of 9, and a deblending contrast ratio `{\tt deblend\_cont}' and a number of deblending thresholds `{\tt deblend\_ntresh}' of 0.001 and 32, respectively). Matched-aperture photometry on the available NIRISS filters is performed at the same stage. From this NIRISS imaging catalogue, the position of sources are used to locate spectral traces in the grism data, and the spectral continua of the sources are modelled using an iterative polynomial fitting of the data for contamination estimate and removal.
These are the NIRISS direct and grism images and contamination models we use in the rest of our analysis.

\subsubsection{Coverage of spectral features}\label{sec:line_coverage}

\begin{figure}
	\includegraphics[width=\columnwidth]{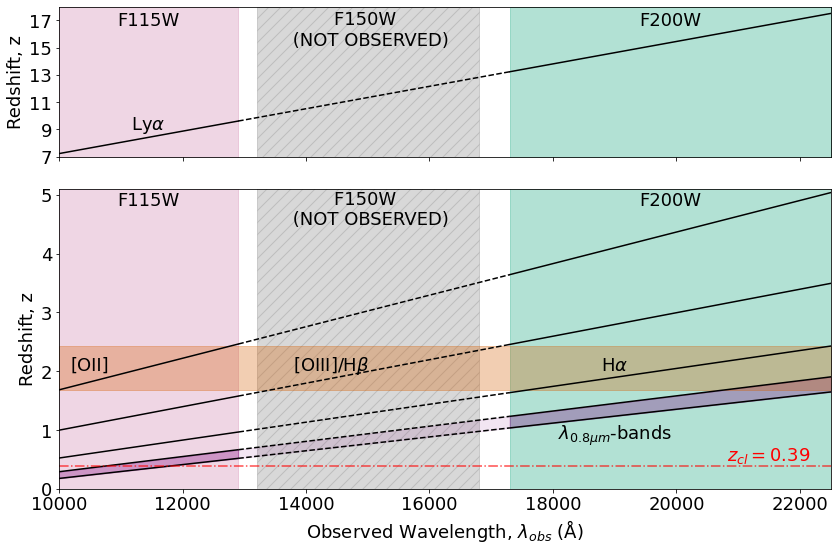}
    \caption{
    Typical strong emission lines (Ly$\alpha$, [OII]$\lambda3727$, [OIII]$\lambda\lambda 4959,5007$, and H$\alpha$, black solid lines) visible in the F115W and F200W grism observations of SMACS~J0723.3$-$7327, shown as a function of redshift and observed wavelength. The diagonal purple band represents the strong molecular bands ($\lambda_{\rm rest}\sim0.8~\micron$) used to identify the cluster galaxies in our NIRISS data, and the red dotted-dashed line indicates the redshift of the cluster ($z_{\rm cl}=0.39$). The light purple and light green vertical shaded areas represent the wavelength coverage of the F115W ($\lambda=1.0-1.3$~\micron) and F200W ($\lambda=1.75-2.25$~\micron) filters, respectively. The vertical grey-hatched shaded area indicates the wavelength gap in the grism coverage of the SMACS~J0723.3$-$7327 EROs. In this gap emission lines cannot be seen, and are indicated with dashed lines. The horizontal light-red shaded area highlights the only redshift range ($z=1.68-2.43$) in this data where at least two typically strong emission lines ([OII]$\lambda3727$ and H$\alpha$) are visible in the F115W and F200W filters at the same time. In this range, the blended [OIII]$\lambda\lambda 4959,5007$ doublet (as well as H$\beta$) falls in the missing F150W filter gap and cannot be detected.
    }
    \label{fig:line_coverage}
\end{figure}

The combination of the F115W and F200W filters ($\lambda=1.0-1.3$ and $1.75-2.25$~\micron, respectively) can in principle allow us to detect H$\alpha$ emission at $z=0.5-1.0$ and $z=1.7-2.4$, [OIII]$\lambda\lambda 4959,5007$ at $z=1.0-1.6$ and $z=2.5-3.5$, [OII]$\lambda3727$ at $z=1.7-2.5$ and $z=3.7-5.0$, and Ly$\alpha$ at $z=7.2-9.7$ and $z=13.4-17.5$.
Figure~\ref{fig:line_coverage} shows the wavelength coverage of the NIRISS WFSS observations together with the emission line visibility as function of redshift within the filters. As indicated in the Figure by the horizontal shaded area, the combination of the F115W and F200W filters, without the intermediate F150W filter ($\lambda=1.32-1.68$~\micron), restricts the redshift range where at least two typically strong emission lines are visible at the same time to $z=1.68-2.43$. Within that range, only [OII]$\lambda3727$ and H$\alpha$ are visible, while the [OIII]$\lambda\lambda 4959,5007$ doublet (blended at the resolution of the grisms; hereafter referred to as [OIII]$\lambda5007$) falls in the missing F150W filter gap. All other redshift ranges have only one (or less) typically strong emission line (among Ly$\alpha$, [OII]$\lambda3727$, [OIII]$\lambda5007$, or H$\alpha$) visible in the F115W and F200W filters. This complicates the identification of emission lines outside of $z\sim1.7-2.4$, and potentially biases our results towards $z\sim2$ sources. Outside of this redshift range of $z=1.68-2.43$, redshift identification with the grism data typically relies on a single strong spectral feature or additional, typically weaker, emission lines. Note that strong Balmer and D4000 absorption features can also be observed in a similar range of redshifts and observed wavelengths as the [OII]$\lambda3727$ emission line shown in Fig.~\ref{fig:line_coverage}. We also use them in our procedure to determine the redshift of the sources when these features are well detected and unambiguously identified.

The H$\alpha$ emission line only enters the F115W filter at redshifts $z>0.52$, which hinders the secure identification and confirmation of the SMACS~J0723.3$-$7327 cluster members (based on emission lines) with the grism data alone. However, we can use the $\lambda_{\rm rest} \sim 0.8~\micron$ absorption features seen in the spectra of bright quiescent galaxies to secure their redshifts. We show the visibility of these features as a light-purple shaded area in the Figure. At the redshift of the cluster ($z_{\rm cl} = 0.39$), these absorption features (primarily due to the presence of TiO in the atmospheres of low-mass main sequence stars -- see, e.g.,  \citealt{GrayCorballyBook}) fall in the F115W filter and can be successfully modelled and fitted as discussed in Sec.~\ref{sec:methods-grism-photoz}.

In principle, the asymmetry of the blended [OIII]$\lambda5007$ doublet can be used to securely determine the redshift of sources in the range $z=1.0-1.6$ and $z=2.5-3.5$. However, we conservatively do not rely on this asymmetry to identify the line in our current work. To determine secure redshifts, we only rely on the clear detection of multiple lines or features (including [OIII]$\lambda5007$ + H$\beta$ only when clearly detected and non-blended), or of a single line whose nature is unambiguously determined based on the photometry (see our methodology in Sec.~\ref{sec:methods-grism-photoz}).

\subsection{NIRspec and MUSE spectroscopic redshifts}
\label{sec:ancillary_specz}

The SMACS~J0723.3$-$7327 field was observed with the Near-InfraRed Spectrograph (NIRSpec) as part of the ERO program \citep{Pontoppidan2022} and a number of papers have reported redshifts based on these observations \citep[e.g.,][]{ Schaerer2022,Carnall2022}.  We here re-analyse the NIRSpec ERO data and recover these redshifts along with eight additional, previously unpublished redshifts, among which four are identified as secure redshifts (spectroscopic $\rm flag=3$) and two as probable redshifts (spectroscopic $\rm flag=2$). This analysis and the resulting redshift list are presented in Appendix~\ref{sec:methods-nirspecredshifts}. 

To complement our dataset, we also use the redshifts published in \citet{Caminha2022} from VLT/MUSE integral field observations of the central $\sim$1~arcmin$^2$ region of the SMACS~J0723.3$-$7327 field.  These MUSE observations are comprised of $50$ secure galaxy redshifts (spectroscopic $\rm flag = 2$ or $3$), nearly half of these being of galaxies in the $z_{\rm cl}=0.39$ cluster, with the rest at higher redshifts up to $z=1.5$.

\section{Methods} \label{sec:methods}

This section describes the generation of the various data products (grism redshift fits, photometry, photometric redshifts) and of our overall methodology to derive our secure spectroscopic catalogue in the field of SMACS~J0723.3$-$7327.

\begin{figure*}
    \centering
    \fbox{\includegraphics[width=1\textwidth]{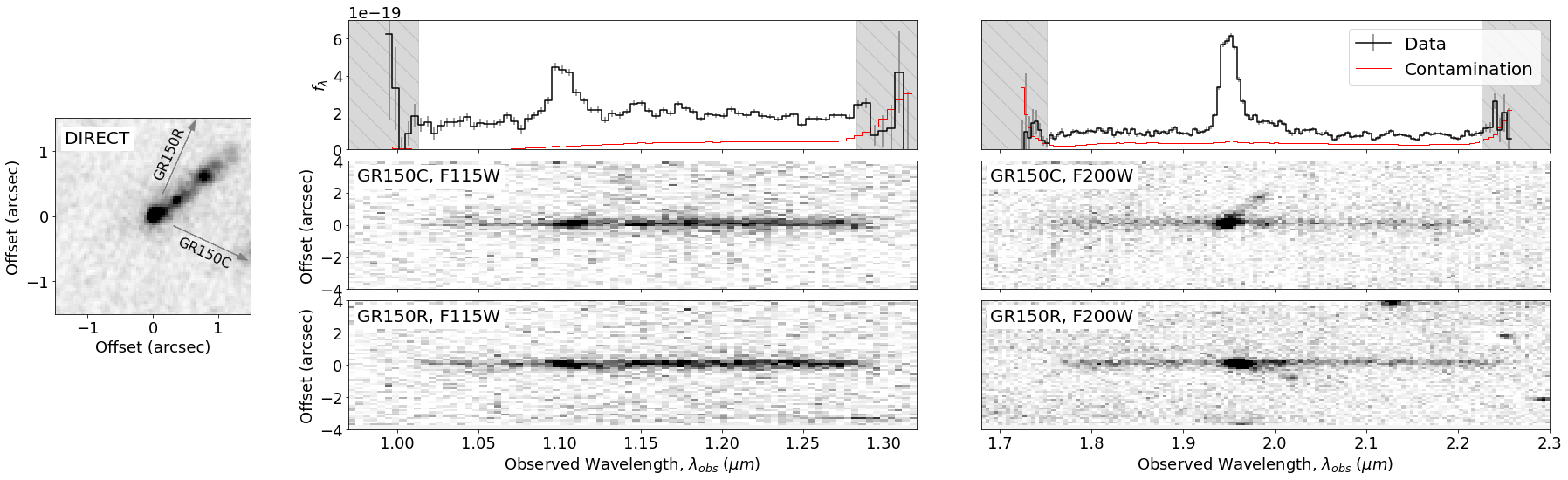}}
    \fbox{\includegraphics[width=1\textwidth]{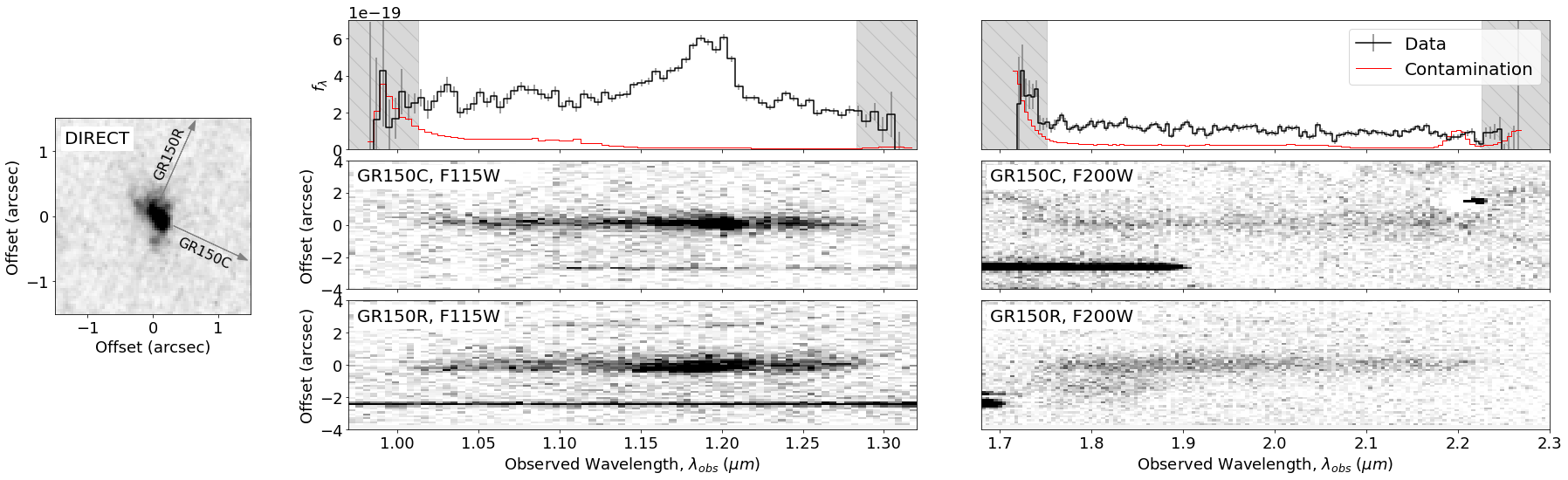}}
    \caption{Examples of quality 1 (top) and quality 2 (bottom) grism spectra. Top panel: example of a quality 1 grism spectrum where multiple emission-lines are clearly identified in the data. In this example, we identify the [OII]$\lambda 3727$ and H$\alpha$ lines at $z=1.97$. Bottom panel: example of a quality 2 grism spectrum where a single emission-line is detected in the data and whose identification is determined based on the photometric redshift PDFs. In this example, we identify the blended [OIII]$\lambda\lambda 4959,5007$ doublet (here also blended with H$\beta$ due to morphological broadening) at $z=1.38$. Each panel shows the direct image stamp of the source (left), its 1D grism spectrum in the F115W and F200W filters (stacked from the two orthogonal grisms; top), and its 2D spectrum in each combination of grism and filter (GR150C, middle, and GR150R, bottom). The dispersion direction of each grism is indicated by the grey arrows overlaid on the direct image stamps. Grey hatched areas in the 1D spectra are regions where the filter transmissions drop below $50\%$ of their peak values. The red solid lines on the 1D spectra are the {\tt grizli} contamination estimates, and the black solid lines with grey error bars are the extracted 1D spectra where contamination is removed.}
    \label{fig:grism_spectra_examples}
\end{figure*}

\begin{figure}
    \centering
    \includegraphics[width=0.48\textwidth]{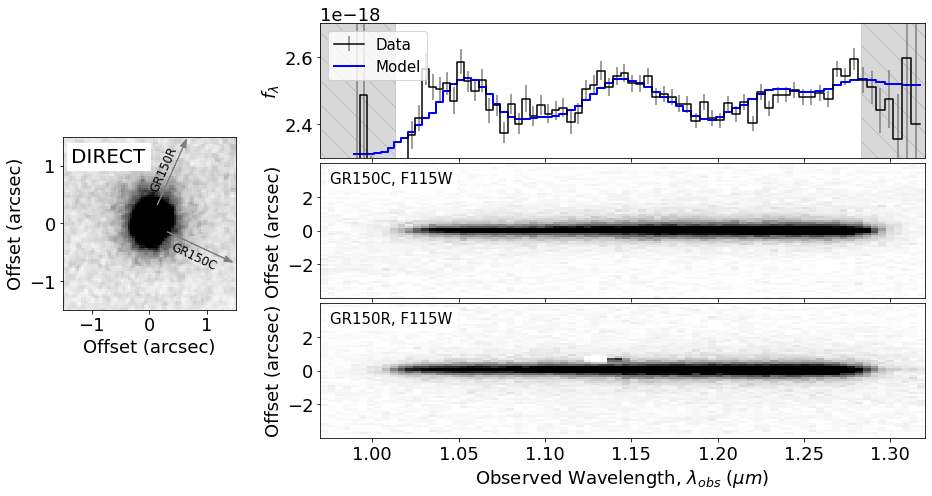}
    \caption{Example of a $z_{\rm cl}=0.39$ cluster galaxy grism spectrum. The panels are the same as in Fig.~\ref{fig:grism_spectra_examples}, but here shown for the F115W filter only. In this example, the spectrum shows the $\lambda_{\rm rest} \sim 0.7-0.9 \micron$ molecular absorption bands (primarily due to the presence of TiO in the atmospheres of low-mass main sequence stars -- see, e.g.,  \citealt{GrayCorballyBook}) that we use to obtain redshifts for low-redshift quiescent galaxies. In the top panel, the blue line is the best-fitted model, which well captures the TiO $\lambda_{\rm rest}\approx 7850$\AA\ and 8600\AA\ absorption features seen at $\lambda_{\rm obs} \sim 1.1$ and $1.2~\micron$.}
    \label{fig:grism_spectra_clg}
\end{figure}

\subsection{Grism redshift fitting}
\label{sec:methods-grismredshifts}

After the reduction steps described in Sec.~\ref{sec:grism_reduction}, we use {\tt Grizli} to extract the 2D grism data of all sources detected in the field and perform an initial redshift fitting of the sources.
For the redshift fitting, we use the default suite of FSPS \citep[Flexible Stellar Population Synthesis;][]{Conroy2009, Conroy2010} templates implemented in {\tt Grizli} as well as {\tt Grizli}'s default emission-line templates and complexes. The advantage of using {\tt Grizli} is that it forward-models the templates on to the 2D grism frames using the source morphology to account for proper spectral smearing of adjacent pixels in the direct imaging (the so-called `self-contamination' or `morphological broadening' of grism data). {\tt Grizli} also removes the contamination from neighbouring objects when fitting the data. We fit the grism data (fitted, with {\tt Grizli}, using the information from the two orthogonal grisms simultaneously) using a redshift range of $z=0-15$, and use the non-negative least square fitting implemented in {\tt Grizli} for the first, coarse fitting within the full redshift range, and the `bounded' fitting (enforcing non-negative continuum templates, but allowing line templates to be negative) for the second, fine fitting around the peaks of the coarse fitting. Then, we use the products of the fitting procedure (redshift PDFs, emission-line maps, best-fitting models), the 2D grism data in each combination of grism and filter, and our photometric redshift measurements (see Section~\ref{sec:photo_z}) to evaluate the quality of the fitting and create our spectroscopic catalogue following our methodology that we describe in Sec.~\ref{sec:methods-grism-photoz}.

Figure~\ref{fig:grism_spectra_examples} shows two examples of grism spectra in our dataset. In the top panel of the Figure, we show an example of a multiple emission-line source (quality 1), and in the bottom panel of the Figure, we show an example of a single emission-line source (whose line is securely identified using our photometric redshift measurements, see our full procedure in Sec.~\ref{sec:methods-grism-photoz}; this is a quality 2 source). Figure~\ref{fig:grism_spectra_clg} shows an example of grism spectrum for a cluster galaxy in our dataset. As already mentioned, cluster galaxies cannot be identified based on the presence of emission-lines in their spectra. However, bright cluster quiescent galaxies can be identified based on the presence of $\lambda_{\rm rest} \sim 0.8~\micron$ absorption features (primarily TiO absorption bands) falling in the F115W filter at the redshift of the cluster. The 1D spectrum in Figure~\ref{fig:grism_spectra_clg} (top row) shows the clear features observed at $\lambda_{\rm obs} \sim 1.1$ and $\sim 1.2~\micron$ and the ability to recover and fit them through our fitting procedure (see Section~\ref{sec:methods-grism-photoz} for our detailed procedure for fitting cluster galaxies).

\subsection{Photometry}
\label{sec:photometry}

A photometric catalogue is created using the imaging data from \textit{HST}/ACS (F435W, F606W, and F814W) and \textit{JWST}/NIRCam (F090W, F150W, F200W, F277W, F356W, and F444W). To detect all the sources in the imaging data, including potential dropouts, we first create a detection $\chi_{\rm mean}$ image (see Sec.~\ref{subsec:detection_image}). The photometry in all of the ACS and NIRCam bands is then measured for all sources detected in the $\chi_{\rm mean}$ image (see Sec.~\ref{sec:photometry-detection-measurement}).

\subsubsection{Detection image}\label{subsec:detection_image}

First, a $\chi_{\rm mean}$ detection image \citep{Drlica-Wagner+2018} is created by combining our {\it HST}/ACS and {\it JWST}/NIRCam images (Sec.~\ref{sec:nircam_hst_imaging}). In this procedure, we use the images before the PSF homogenization step.
For each pixel, the $\chi_{\rm mean}$ signal is computed as
\begin{equation}
\label{eq:chi_mean}
\chi_{\rm mean} = \frac{\sqrt{\sum\limits_{i\le n} {f_i^2}/{\sigma_i^2}} - \mu} {\sqrt{n - \mu^2}},
\end{equation}
where  $f_i$ is the flux in the $i^{\rm th}$ band, $\sigma_i^2$ its associated variance, $n$ is the number of input images used in the combination and 
$$\mu = \sqrt{2}\frac{\Gamma((n+1)/2)}{\Gamma(n/2)}.$$
The inclusion of the offset term $\mu$ in Eq.~\ref{eq:chi_mean} results in the suppression of artifacts at the edges of individual input images in the $\chi_{\rm mean}$ stack (see \citealt{Drlica-Wagner+2018}) and is important in reducing the number of false object detections.

Ideally, the variance maps in the different bands must represent the true variance in the images, which can often not be the case due to image processing (e.g., due to drizzling). Thus, we here apply a re-scaling of the variance map to ensure that proper noise is propagated to the detection image. It is done by comparing the root-mean-square (RMS) noise in noise-normalized images to the RMS computed from variance maps. We implement this process by dividing the images in three regions to account for the difference of noise level due to image stacking.

\subsubsection{Object detection and photometry}
\label{sec:photometry-detection-measurement}
We use version 1.5.0 of the \photutils \citep{larry_bradley_2022_6825092} package to perform object detection and photometry.
We run object detection on the $\chi_{\rm mean}$ detection image, and use a two-mode ``hot + cold'' detection strategy which has been shown in previous surveys to be successful and effective in providing reliable detections for both bright, large and faint, small objects \citep[e.g., CANDELS;][]{guo_candels_2013, Galametz2013}. The cold mode detection parameters are set less aggressively so that bright and large objects can be detected without being overdeblended, while the hot mode parameters are set more aggressively to detect fainter and smaller objects. We then follow the strategy of \citet{barden_galapagos_2012} for combining the outputs of the two modes: all objects from the cold mode detection are kept in the final catalogue, but objects from the hot mode detection located within any of the Kron apertures of cold-mode objects are regarded as parts of substructures of cold-mode objects and are removed from the final catalogue.  We run this two-mode detection on our $\chi_{\rm mean}$ detection image of the SMACS~J0723.3$-$7327 field and detect a total of $9703$ objects.

For each of the detected objects, photometry measurements in all images are performed with the same strategy as Source Extractor dual-image mode \citep{Bertin1996}. We use the $\chi_{\rm mean}$ detection image to determine the Kron aperture for each detected object, and measure fluxes within the aperture on the F444W PSF-matched images to obtain the Kron flux in each band. Additionally, we also perform circular aperture photometry with the diameter of $0\farcs7$ at the position of each object. The photometry is then corrected for Galactic extinction using the value of $E(B-V)=0.19$~mag (AB) from \citet{schlafly_measuring_2011} for the right ascension and declination of the SMACS~J0723.3$-$7327 field and assuming the \citet{fitzpatrick_correcting_1999} extinction law, with $R_V=3.1$. The photometric errors are estimated from the re-scaled variance maps (see Sec.~\ref{subsec:detection_image} for details) to take in account the underestimation of the errors in the weight maps due to image processing and drizzling.

\subsection{Photometric redshifts}\label{sec:photo_z}

To securely identify the nature of single emission-lines in the grism data (see our methodology in Sec.~\ref{sec:methods-grism-photoz}), we use photometric redshift estimates that we derive from our photometric catalogue. For this, we choose to run two template-fitting codes with different configurations, namely \Phosphoros (\citealt{Desprez2020}; Paltani et al., in prep) and \texttt{EAzY} \citep{Brammer2008}. This strategy allows us to mitigate against errors and systematics, both in the (still novel) {\it JWST} photometry and in the photometric redshift fits themselves (especially given the relative lack of spectroscopic redshifts available ahead of our work to properly assess the quality of our photometric redshifts). 

\subsubsection{Phosphoros photometric redshifts}

For the photometric redshifts computed using {\tt Phosphoros}, the 33 spectral energy distributions (SEDs) from the \citet{Ilbert2013} template library are used, including seven elliptical-galaxy SEDs, twelve spiral-galaxy SEDs \citep{Polletta2007,Ilbert2009}, twelve starburst-galaxy SEDs and two elliptical-galaxy SEDs generated with the \citet{BruzualCharlot2003} stellar population synthesis models. Intrinsic extinction is set as a free parameter with reddening excess ranging from $0$ to $0.5$~mag (AB), and considering different attenuation laws: ({\it i}) the Calzetti law \citep{Calzetti2000} for templates bluer than the Sb template, ({\it ii}) 
two versions of the Calzetti law including the 2175\AA\ bump for templates bluer than SB03, ({\it iii}) a Small-Magellanic-Cloud-like law \citep{Prevot1984} for templates redder than SB3, and ({\it iv}) no extinction for Sb templates and redder.

As described in Paltani et al.~(in prep.), \Phosphoros includes emission lines in the templates by assuming an empirical relation between the UV continuum and H${\alpha}$ line fluxes \citep{Kennicutt1998}, and ratios between the different hydrogen lines and oxygen lines as measured in the Sloan Digital Sky Survey (SDSS; \citealt{York+2000}) galaxy spectra. 
No magnitude prior is applied. However, a prior is applied on the SEDs, weighing the SEDs according to their overlap in rest-frame color-space, thus mitigating the risk to overweight solutions produced by too similar templates (Paltani et al., in prep).
The redshift fits are performed on the $0\farcs7$-diameter aperture photometry measured in the PSF-homogenized images {and already corrected for Galactic extinction}. Photometric redshift probability distribution functions (PDFs) are computed by marginalizing the posterior grid obtained through $\chi^2$ minimization and the application of the priors. The PDFs are defined on a $z=0$--$16$ grid with a redshift step of $\delta z=0.02$.

\subsubsection{EAzY photometric redshifts}

We also use {\tt EAzY} to compute photometric redshifts for all sources with a configuration close to the default one. The default SEDs templates (i.e., {\tt tweak\_fsps\_QSF\_v12\_v3}) are used to fit the Kron photometry, allowing for multiple template combinations and emission line additions. No additional zero-point correction is applied to the photometry at this stage, but a systematic error of 3\% of the flux is added in quadrature to the photometric uncertainties in each band.
A magnitude prior in the form of a $p(z|m_{\rm F150W})$ is also used. The PDFs are sampled on a redshift grid within $z=0 - 12$ with a step of $\delta z=0.01$ after the application of the prior.

\subsection{Building the grism redshift catalogue}
\label{sec:methods-grism-photoz}

\begin{figure*}
    \centering
    \includegraphics
    [width=0.65\textwidth]
    {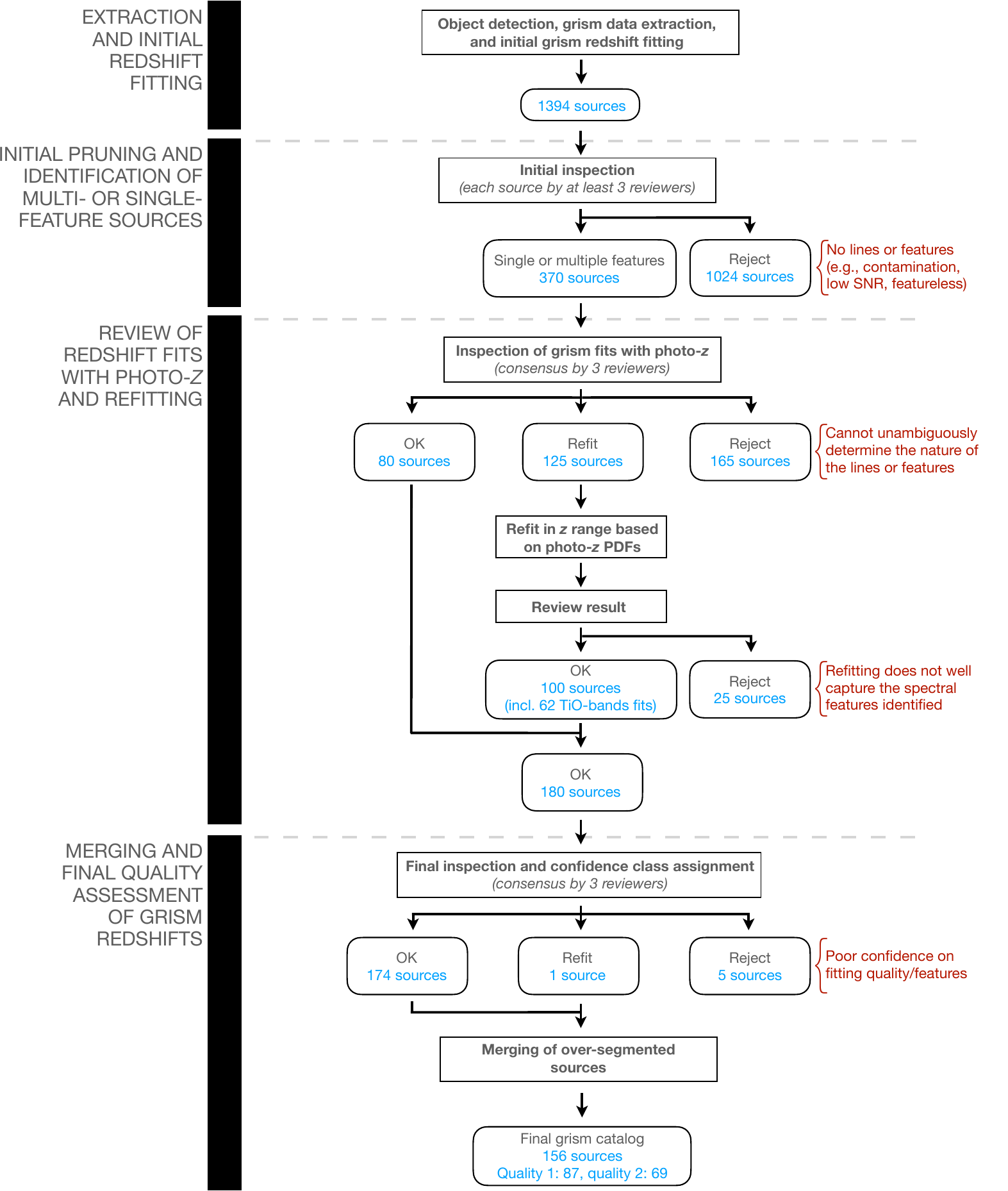}
    \caption{Flowchart of the methodology we employ to obtain our final sample of $156$ secure NIRISS grism redshifts. Our methodology is comprised of four main stages: ({\it i}) initial extraction and redshift fitting of all grism sources (1394 sources), ({\it ii}) initial pruning and identification of sources with single or multiple spectral features (370 sources), ({\it iii}) review of the grism redshift fits aided by the {\tt EAzY} and {\tt Phosphoros} photometric redshift PDFs (180 sources), and ({\it iv}) final quality assessment of the grism redshifts and merging with the photometric source catalogue (156 sources). The red-coloured text on the right-hand side of the flowchart describes the reasons for discarding sources in the different inspection stages of our methodology. This sample is then complemented with the NIRSpec and MUSE data described in Sec.~\ref{sec:ancillary_specz}. Together, this constitutes our final catalogue of $190$ secure spectroscopic redshifts in the field of SMACS~J0723.3$-$7327. See Sec.~\ref{sec:methods-grism-photoz} for a detailed description of the flowchart.}
    \label{fig:flowchart}
\end{figure*}

Figure~\ref{fig:flowchart} shows the flowchart of the methodology we employ to obtain our final sample of secure NIRISS grism redshifts. As described in Section~\ref{sec:grism_reduction}, we first use {\tt Grizli} to detect objects and extract their grism data in the NIRISS observations, and also to perform an initial redshift fitting of the grism data in the range $z=0-15$ for all detected objects (Sec.~\ref{sec:methods-grismredshifts}). This gives us an initial sample of $1394$~NIRISS sources each with a {\tt Grizli} redshift PDF and associated data products (e.g., emission-line maps, best-fit model, rest-frame EWs, etc). 

After this initial step, we visually inspect the spectral quality of the grism data and reject sources which do not have clear or tentative spectral features (single or multiple emission-lines, absorption features, and spectral breaks). After an initial training on a sample of $100$~sources, this step is performed by six reviewers (GN, GD, MS, VEC, YA, SM) on different subsets of the sample, with each source being inspected by a minimum of three reviewers. From this first visual inspection, all sources for which $>50\%$ of the three or more reviewers assign the lowest spectral quality (no clear or tentative spectral feature) are rejected from the sample. Typically, these sources are rejected due to ({\it i}) the imperfect subtraction of contamination from neighbouring sources, ({\it ii}) low signal-to-noise ratio (SNR) data, or ({\it iii}) featureless spectral continua. Due to the high density of sources in the field and the relatively shallow depth of the NIRISS observations, this step rejects $1024$~sources ($73\%$), while $370$~sources ($27\%$) with clear or tentative spectral features are kept in the sample for further assessment. 

In the third stage of our methodology, three reviewers (GN, GD, MS) visually inspect the results of the grism redshift fitting for the $370$~sources kept in the sample. The reviewers first assess the quality of the results independently, and then review discrepant assessments together to reach a consensus on the fitting quality. For this inspection, the {\tt Grizli} grism redshift fits are compared to the two photometric redshift PDFs obtained with \texttt{EAzY} and \texttt{Phosphoros}. Sources are assigned an ``OK'' flag if ({\it i}) multiple features are unambiguously identified in their grism data and are well fitted (i.e., no strong contamination in one or multiple grism/filters and both grism orientations show consistent features), or ({\it ii}) a single feature is unambiguously identified in the grism data and the nature of the feature is securely determined and consistent between the {\tt Grizli} redshift solution and the two solutions from the photometric redshift PDFs. Sources with clear single features but inconsistent solutions between the {\tt Grizli} and photometric redshifts are selected for refiting of their grism data with {\tt Grizli} in a narrower redshift range. This is done only if the photometric redshift PDFs allow to unambiguously determine the nature of the feature detected in the grism data. In those cases, the grism data is refitted within $\pm0.15\times(1+z)$ of the redshift solution that is consistent with the position of the grism feature. However, if ({\it i}) the photometric redshift PDFs are consistent with solutions around the $z_{\rm cl}=0.39$ cluster redshift, and ({\it ii}) the spectral features in the grism data are indicative of the $\lambda_{\rm rest}\sim0.8~\micron$ molecular absorption bands (primarily due to the presence of TiO in the atmospheres of low-mass main sequence stars -- see, e.g.,  \citealt{GrayCorballyBook}) redshifted to $z\sim0.4$ (see Fig.~\ref{fig:grism_spectra_clg}), we use a custom redshift range of $z=0.33-0.45$ to refit the grism data. Sources for which the nature of the single spectral features cannot be unambiguously determined based on the photometric redshifts are discarded. In this third stage of inspection, we reject a total of $165$~sources ($45\%$) with ambiguous spectral features, refit $125$ ($33\%$), and keep $80$ ($22\%$) without further refitting.

As mentioned, among the $125$~sources flagged for refiting, sources consistent with being at the cluster redshift are refitted within $z=0.33-0.45$. To better capture the molecular absorption bands seen in the grism data of those potential cluster galaxies, we additionally complement the default set of {\tt Grizli} templates with $18$ new FSPS quiescent galaxy templates spanning a range of ages and metallicities. We create this additional template set using single stellar population (SSP) star-formation histories (SFHs), a \citet{Kroupa2001} initial mass function (IMF), the \citet{Calzetti2000} dust law with fixed attenuation of $A_V=0.2$~mag (AB), Cloudy photoionization for nebular emission \citep{Ferland2017}, and \citet{Madau1995} intergalactic medium (IGM) absorption. The FSPS models also include the \citet{Draine2007} dust emission models, and the \citet{Villaume2015} AGB circumstellar dust models. Note that while we use IGM absorption and nebular emission in our models, their effects in our new set of template should be relatively minor as we are generating old, quiescent templates for which the wavelength range of interest in our work is primarily the near infrared. We use two sets of ages and metallicities to create these additional quiescent templates. We create 9 templates using combinations of ${\rm Ages} = [6,9,10]$~Gyrs and $Z =[0.5,1,1.5]~Z_{\odot}$, and another set of 9 templates with ${\rm Ages} = [9,11,13]$~Gyrs and $Z = [2,2.5,3.0]~Z_{\odot}$.
When fitting the cluster galaxies, we also remove the emission-line templates and complexes from the default {\tt Grizli} fitting routine. In a number of cases, we refit the F115W grism data alone (as this contains the redshifted $\lambda_{\rm rest}\sim0.8~\micron$ molecular absorption features) to better capture the grism absorption features when the fits are perturbed by the F200W data.

After refitting the $125$~sources, the results are again reviewed by the three reviewers (GN, GD, MS) and either validated or rejected as in the previous step. The rejected sources are sources for which the refitting does not well capture the spectral features identified in the data. At this stage, we reject $25$ of the $125$~sources, and keep $100$ (including $62$ fitted using the TiO absorption bands) in our sample. Together with the 80~sources previously flagged as "OK", this gives us a sample of $180$~sources flagged with ``OK'' grism-fit quality ready for a final inspection.

In the final step of our procedure, the three reviewers (GN, GD, MS) visually inspect the grism redshift fits of the $180$~sources with ``OK'' grism-fit quality for final vetting and quality assessment. In this step, five sources are rejected from the final catalogue due to poor confidence on the spectral features, and one is refitted (within our initial redshift range of $z=0-15$) using a slightly narrower spectral coverage in the F115W and F200W filters than the original fit of this source to remove edge effects biasing the fit. After this procedure, $175$~sources pass our final vetting and $90$ ($51\%$) are given a quality 1 class (i.e., multiple features), and $85$ ($49\%$) a quality 2 class (i.e., single features supported by photo-$z$). This conservative sample of $175$ secure grism redshifts represents $13\%$ (175/1394) of the original sample. 

Last, we identify the grism objects in our initial {\tt Grizli} extraction that correspond to single sources in the photometric catalogue, and we merge those multi-component grism sources in our final catalogue. Specifically, when multiple grism sources fall within the segmentation region of a single photometric object, we verify that the grism redshifts associated with the photometric objects are consistent with one another and report the average grism redshift of the components as the source redshift. For those sources, the quality flag is reported as the highest quality flag from among the multiple components. This step merges together $34$ grism sources to $15$ unique photometric objects. This gives us a total of $156$ secure NIRISS grism redshifts, including $87$ ($56\%$) quality 1 and $69$ ($44\%$) quality 2 sources. 

These 156 high-quality NIRISS redshifts form the basis of our catalogue.
As mentioned in Sec.~\ref{sec:ancillary_specz}, we additionally complement the catalogue with the secure NIRSpec (see Appendix~\ref{sec:methods-nirspecredshifts}) and MUSE data (spectroscopic $\rm flag=2$ and $\rm flag=3$) available in the field (either published in the literature or from our own analysis). This complements our catalogue with an additional $65$ spectroscopic redshifts, including $31$ for which we have an independent grism redshift measurement in our NIRISS redshift sample. Overall, our final catalogue of secure NIRISS, NIRSpec and MUSE spectroscopic redshifts contains a total of $190$ sources, including 156 NIRISS grism redshifts, $123$ of which are for sources whose redshifts were previously unknown. Table~\ref{tab:catalog} lists the contents of our final spectroscopic catalogue of $190$ sources, while Table~\ref{tab:catalog2} shows the detail of each component for the multi-component sources. The two catalogues are made available at \url{https://niriss.github.io/smacs0723}.

\section{Properties of the redshift sample}\label{sec:redshift_sample}

\subsection{Redshift fidelity}\label{sec:redshift_fidelity}
  
\subsubsection{Spectroscopic versus grism redshifts}

\begin{figure}
    \centering
    \includegraphics
    [width=0.45\textwidth]
    {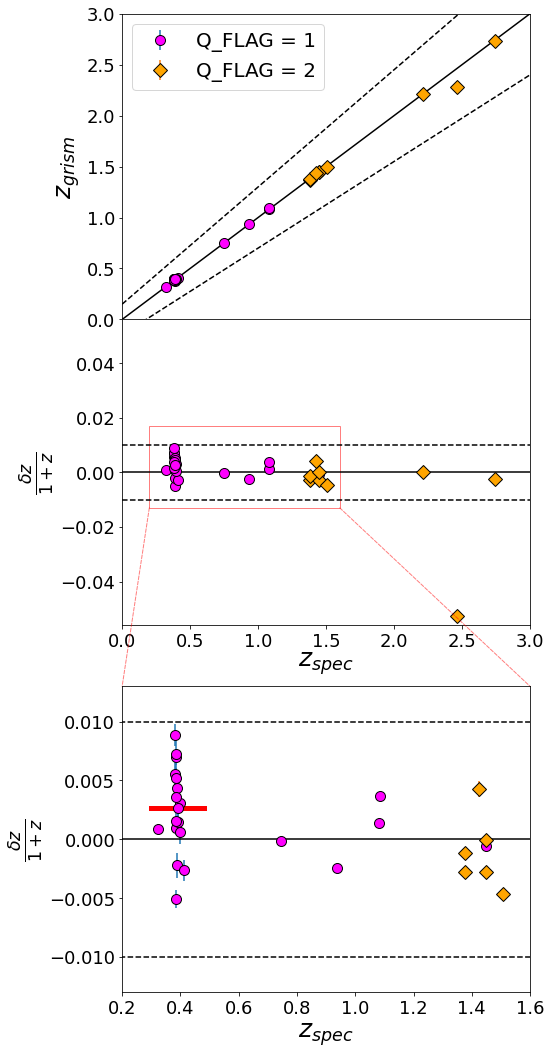}  
    \caption{Grism redshift fidelity as a function of spectroscopic redshift. In all panels, the magenta circles indicate sources with multiple spectral features ($\rm Q\_FLAG = 1$), and orange diamonds indicate sources with single spectral features aided by photometry ($\rm Q\_FLAG = 2$). The top panel shows $z_{\rm grism}$ vs.~$z_{\rm spec}$. The solid line is the 1:1 relation, and the dashed lines show the $\pm 0.15 \times (1+z)$ deviation around the line. The middle panel shows the redshift deviation, $\frac{\delta z}{1+z_{\rm spec}}$, with $\delta z = z_{\rm grism} - z_{\rm spec}$, as a function of spectroscopic redshift. In this panel, the horizontal dashed lines show $\pm 0.01$ level deviation. The bottom panel shows a zoom-in view of the $\frac{\delta z}{1+z_{\rm spec}}=\pm0.01$ interval within $z_{\rm spec}=0.2-1.6$. Cluster galaxies show a small systematic bias of the order of $\delta z = 2.6 \cdot 10^{-3} \times (1+z)$, indicated by the thick red line. In most cases, the size of the error bars {(random uncertainties)} in the different panels are smaller than the data points.}
    \label{fig:zgrism_vs_zspec}
\end{figure}

Our final catalogue includes $65$ NIRSpec and MUSE spectroscopic redshifts. Among those, 31 are sources for which we independently measure a NIRISS grism redshift. We use this sample of 31 sources with independent spectroscopic and NIRISS grism redshift measurements to evaluate the redshift fidelity of our NIRISS grism work. In the top panel of Figure~\ref{fig:zgrism_vs_zspec}, we show the agreement between grism and spectroscopic redshifts (i.e., $z_{\rm grism}$ vs.~$z_{\rm spec}$). The solid line represents the 1:1 relationship, while the dashed lines show the $0.15\times (1+z)$ deviation around the line. We highlight with magenta circles sources that have a grism quality of 1 ($\rm Q\_FLAG = 1$) and with orange diamonds sources that have a grism quality of 2 ($\rm Q\_FLAG = 2$). As seen in the Figure, all sources are within the $0.15\times (1+z)$ deviation, and all but one very closely follow the 1:1 line. This is true for all galaxies whose NIRISS grism redshifts are based on the identification of multiple features, but is also true for all galaxies (except one showing a slight offset) whose NIRISS grism redshifts are based on the identification of a single feature whose nature is securely determined based on the {\tt EAzY} and {\tt Phosphoros} photometric redshift PDFs. This demonstrates the very high fidelity and quality of our NIRISS grism redshifts for both quality 1 and quality 2 sources; {and in particular demonstrates the very high fidelity of our strategy of using photometric redshifts to disentangle between multiple line solutions for quality 2 sources. The latter is in part enabled by the good agreement between our {\tt EAzY} and {\tt Phosophoros} photometric redshifts ($\sigma_{\rm NMAD} < 0.05$ and outlier fraction $\eta < 15\%$), and between our photometric redshifts and the limited sample of NIRSpec and MUSE spectroscopic redshifts ($\sigma_{\rm NMAD} = 0.03$ ($0.04$) and outlier fraction $\eta = 4\%$ ($14\%$) for {\tt EAzY} ({\tt Phosphoros}) photometric redshifts).} {The high fidelity of our NIRISS grism redshifts} is also seen in the middle panel of the Figure, which shows the redshift deviation ($\frac{\delta z}{1+z_{\rm spec}}$, with $\delta z = z_{\rm grism} - z_{\rm spec}$) as a function of spectroscopic redshift. In this panel, all quality 1 and quality 2 sources are within $\pm 0.01 \times (1+z)$ of the spectroscopic redshifts (highlighted by the dashed lines). 

The only source slightly deviating from this threshold is source ID~750 (c.f.~Table~\ref{tab:catalog}), with a quality 2 and a deviation of $\frac{\delta z}{1+z_{\rm spec}}=-0.05$. This source has a NIRSpec redshift of $z_{\rm spec} = 2.463$ \citep{Carnall2022}. At this redshift, [OII]$\lambda3727$, [OIII]$\lambda5007$, and H$\alpha$ are expected to fall at the edges of the wavelength coverage of the F115W and F200W filters, which prevents the measurement of such a redshift based on those emission-lines with the NIRISS grism data. In the NIRISS data of this source, we identify a single emission-line at $\lambda_{\rm obs} = 2.15$~\micron, together with a consistent photometric redshift between {\tt EAzY} and {\tt Phosphoros} of $z_{\rm phot} \sim 2.3\pm0.2$, indicating that the emission-line is consistent with H$\alpha$ at $z_{\rm grism} = 2.28$. While the line identified in the grism data is relatively weak, it is ({\it i}) well detected in the two orthogonal grisms, ({\it ii}) the overall wavelength range in the F115W and F200W show little to no contamination, and ({\it iii}) the line is well aligned with the position of the grism source. This indicates that the emission line is real and associated with the object detected in the {\tt Grizli} grism extraction.
However, the source is relatively extended in the NIRCam imaging, and the centroid position of the photometric object and of the NIRSpec slit are both offset by $\sim 1$~arcsec from the position of the grism object. Specifically, the grism object corresponds to a bright knot offset by $\sim 1$~arcsec to the South of the center of the photometric source. Overall, this suggests that the difference between the grism and spectroscopic redshifts could be due to: ({\it i}) a real chance alignment between a foreground and a background object, or ({\it ii}) contamination or confusion on the nature of the line identified in the grism data. We exclude the possibility of a real rotational velocity offset given the observed redshift difference. While further analysis is necessary to determine the nature of the slight redshift discrepancy for this source, we keep it in the present catalogue and report both its NIRSpec and NIRISS measurements.

In the bottom panel of Figure~\ref{fig:zgrism_vs_zspec}, we show a zoom-in view of the $\frac{\delta z}{1+z_{\rm spec}}=\pm0.01$ interval within $z_{\rm spec}=0.2-1.6$. While the agreement between spectroscopic and NIRISS grism redshifts are excellent for both quality 1 and quality 2 sources, this panel shows the slight systematic offset that may be present in the grism measurement of the cluster galaxy redshifts. For $z<0.5$ sources, we measure a median deviation of $\delta z = 2.6\cdot10^{-3} \times (1+z)$. The level of this deviation is shown as the thick red line over $z_{\rm spec} = 0.30 - 0.48$ in the Figure panel. We do not correct for this potential systematic offset in the present catalogue as it appears quite marginal, but note that it is a potential (marginal) bias in our measurement.

Note that \citet{Li2022} reports seven NIRISS grism redshifts in the SMACS~J0723.3$-$7327 field as part of a larger analysis of the mass-metallicity relation of dwarf galaxies at cosmic noon. Three of these redshifts are independently recovered in our catalogue while the other four are rejected by our screening procedure detailed in Sec.~\ref{sec:methods-grism-photoz}. Appendix~\ref{sec:Li-comparison} discusses the \citet{Li2022} sample in more detail. 

\subsubsection{Field variation}
\label{subsection:field-variation}

\begin{figure}
    \centering
    \includegraphics
    [width=0.45\textwidth]
    {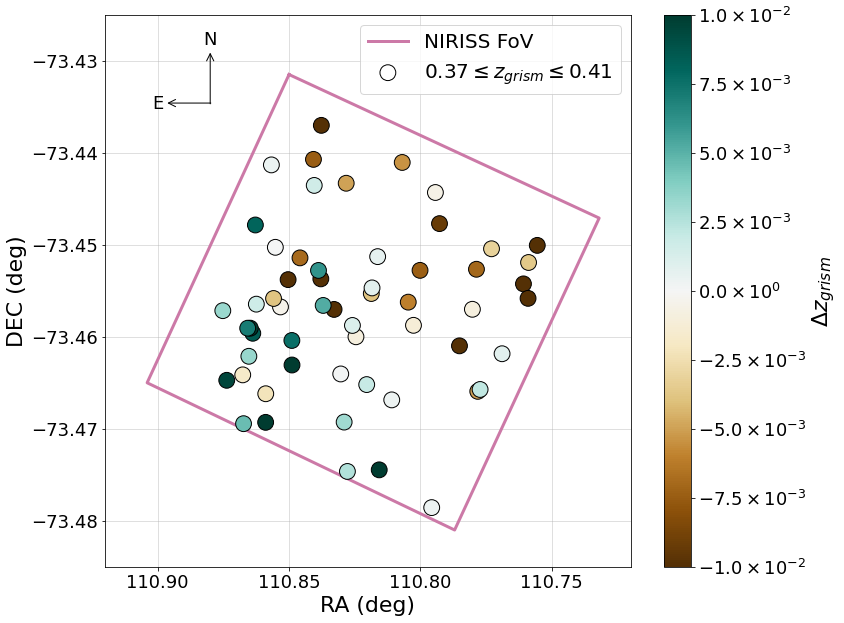}
    \caption{Grism redshift variation across the field for galaxies within $0.37 \leq z_{\rm grism} \leq 0.41$ (coloured circles). The light-magenta contour line represents the NIRISS field of view; North is up and East is to the left. The colour of the data points represents the difference (noted $\Delta z_{\rm grism}$) between the redshift of the sources and $\tilde{z}=0.394$, the median redshift within $0.37 \leq z_{\rm grism} \leq 0.41$. A slight trend of the order of $|{\Delta z_{\rm grism}}| \leq 0.01$ can be seen in the North-South direction. This trend could be due to a real redshift/velocity offset in the $z_{\rm cl} = 0.39$ SMACS~J0723.3$-$7327 cluster across the field or slight field-dependent systematics affecting the NIRISS grism wavelength solutions.}
    \label{fig:field_variation}
\end{figure}

As seen in Figure~\ref{fig:zgrism_vs_zspec}, there is an overall excellent agreement between the spectroscopic and grism redshift measurements of both quality 1 and quality 2 sources in our catalogue. In these early days of science with {\it JWST}, it is however important to assess the different biases that may affect our data, data reduction, and measurements. In Figure~\ref{fig:field_variation}, we show the variation of our grism redshift measurements across the NIRISS field-of-view for galaxies with a measured grism redshift consistent with the redshift of the SMACS~J0723.3$-$7327 cluster. We select all galaxies within $0.37 \leq z_{\rm grism} \leq 0.41$, and show as coloured circles the difference between the redshift of the sources and $\tilde{z}=0.394$, the median redshift of the sources within that range. We denote this difference $\Delta z_{\rm grism}$. 

As seen in the Figure, there seems to be a slight trend of the order of $|{\Delta z_{\rm grism}}| \leq 0.01$ with respect to the North-South direction, with slightly negative $\Delta z_{\rm grism}$ towards the North of the field, and slightly positive $\Delta z_{\rm grism}$ towards the South. This trend could be due to a real redshift/velocity offset in the $z_{\rm cl} = 0.39$ SMACS~J0723.3$-$7327 cluster across the field or due to slight field-dependent systematics affecting the NIRISS grism wavelength solutions in our data. {During the writing of this paper, \citet{grizliconf2022} provided updated wavelength and cross-dispersion zero-points as well as sensitivity curves for NIRISS GR150R and GR150C grisms in the F115W, F150W, and F200W filters. However, using these updated calibrations do not significantly change our results. We therefore do not use these in the current paper.}

Note, however, that {the potential trend seen in the North-South direction} currently appears limited and, if caused by systematics {and not due to a real redshift or velocity offset}, should not affect the overall quality of our grism redshift measurements beyond the $1\%$ level. Also, we do not observe such a trend with respect to field position when comparing our grism and spectroscopic redshifts for both cluster and non-cluster galaxies, {which would support that the trend is not a systematic bias but a real redshift feature.} However, spectroscopic sources are mostly located within the central part of the field which was observed with MUSE. This might prevent us from observing a similar field-dependent variation as seen with our grism sample of cluster galaxies.

\subsection{Redshift distribution}
\label{subsec:redshift-distribution}

\begin{figure*}
    \centering
    \includegraphics
    [width=1\textwidth]
    {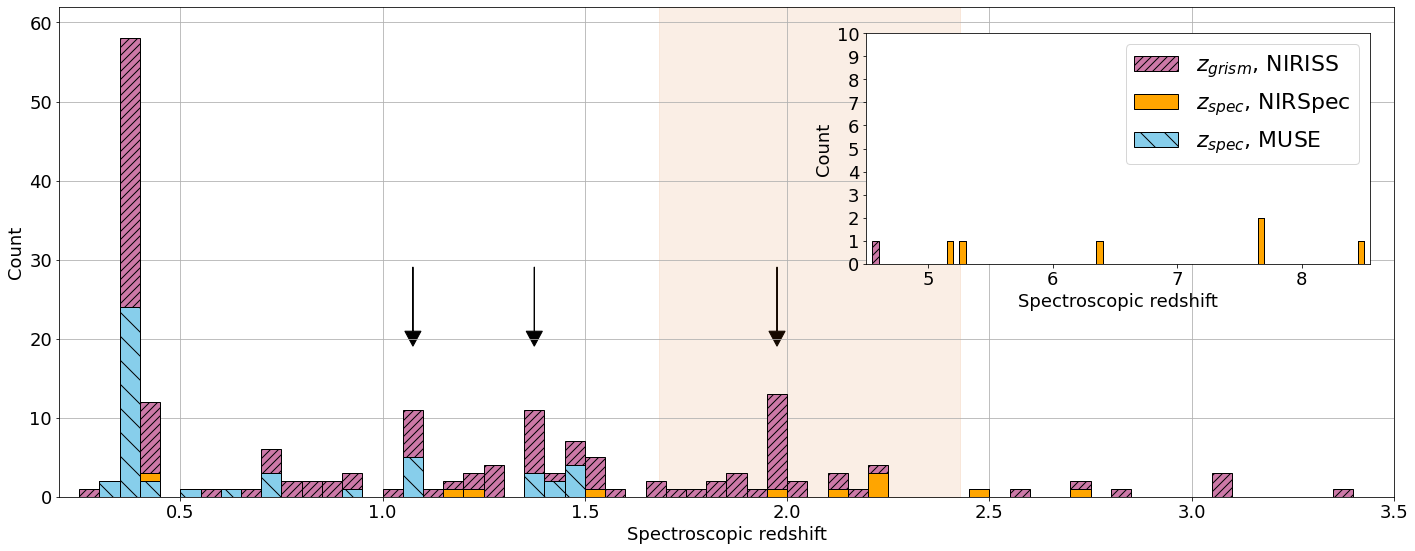}
    \caption{Redshift histogram of our final spectroscopic catalogue (c.f.~Table~\ref{tab:catalog}). The main panel shows the redshift distribution up to $z=3.5$, and the inset highlights the $z = 4.5 - 8.5$ range. NIRSpec and MUSE redshifts are shown in orange and hatched-blue, respectively. The NIRSpec redshifts include the six new redshifts we measure in our work (Sec.~\ref{sec:methods-nirspecredshifts}). We show our new NIRISS redshifts in double-hatched purple. These are the new secure redshifts we obtain in our work which do not have a NIRSpec or MUSE measurement. The histogram shows the clear peak at the cluster redshift ($z_{\rm cl} = 0.39$). Three additional peaks at $z=1.1$, $z=1.4$, and $z=2.0$, are seen in the histogram and highlighted by black arrows. The light-red shaded area shows the redshift range where two typically strong emission-lines are visible at the same time in the NIRISS data of the field (c.f.~Sec.~\ref{sec:line_coverage} and Fig.~\ref{fig:line_coverage}). The narrow redshift peak seen at $z=2.0$ suggests it is real and not an instrumental bias.}
    \label{fig:redshift_histogram}
\end{figure*}

Figure~\ref{fig:redshift_histogram} shows the redshift distribution of our final spectroscopic catalogue (Table~\ref{tab:catalog}). We show NIRSpec and MUSE spectroscopic redshifts in orange and hatched-blue, respectively, while the double-hatched purple indicates the new NIRISS grism redshifts obtained in our analysis that do not have a NIRSpec or MUSE counterpart. The main panel shows the redshift distribution up to $z=3.5$, and the inset highlights the $z=4.5-8.5$ range (the $z=3.5-4.5$ range does not contain any source in our catalogue). As clearly seen in the Figure, our NIRISS analysis recovers the expected peak at the redshift of the SMACS~J0723.3$-$7327 cluster ($z_{\rm cl}=0.39$) and more than doubles its number of spectroscopic members. At redshifts $z>0.5$, our NIRISS (NIRSpec) work identifies 81 (5) new sources, increasing the number of sources behind SMACS~J0723.3$-$7327 from $31$ to $117$ (including the 5 new $z>0.5$ NIRSpec sources).

This significant improvement allows us to identify three galaxy overdensities behind SMACS~J0723.3$-$7327 that were until now missed by the existing {\it JWST}/NIRSpec and VLT/MUSE data of the field. These overdensities, highlighted by the black arrows on the Figure, are at $\overline{z}=1.08$, $1.37$, and $1.98$, with $\gtrsim 10$ secure spectroscopic members each.

\begin{itemize}
\item {\bf $\overline{z}=1.08$ overdensity.} In the first overdensity, we identify 11 sources within $z=1.05-1.09$, including five based on MUSE spectrosocpy and six based solely on our NIRISS grism analysis. One grism source (ID~680 in our final catalogue; c.f.~Table~\ref{tab:catalog}) possesses four grism components within $\sigma_z=\pm0.002$ of $\overline{z}=1.085$ (IDs 680.1, 680.2, 680.3, 680.4 in our multi-component catalogue; c.f.~Table~\ref{tab:catalog2}), while another source (ID~1155) possesses one grism redshift and two MUSE counterparts with consistent redshifts (IDs~1155.1 and 1155.2). The spatial distribution of the 11 sources over the field is presented in the left panel of Figure~\ref{fig:spatial_dist} (red circles). In the right panel of the Figure, we show a close-up view on four of the sources (IDs~1155, 1156, 1157, 1158) which seem to form a closely-interacting system (indicated by the black ellipse on the left panel in Fig.~\ref{fig:spatial_dist}). These four sources have MUSE spectroscopic redshifts of $z_{\rm spec}=1.082$, $1.082$, $1.083$, and $1.083$, respectively.

\begin{figure*}
    \centering
    \includegraphics
    [width=0.49\textwidth]
    {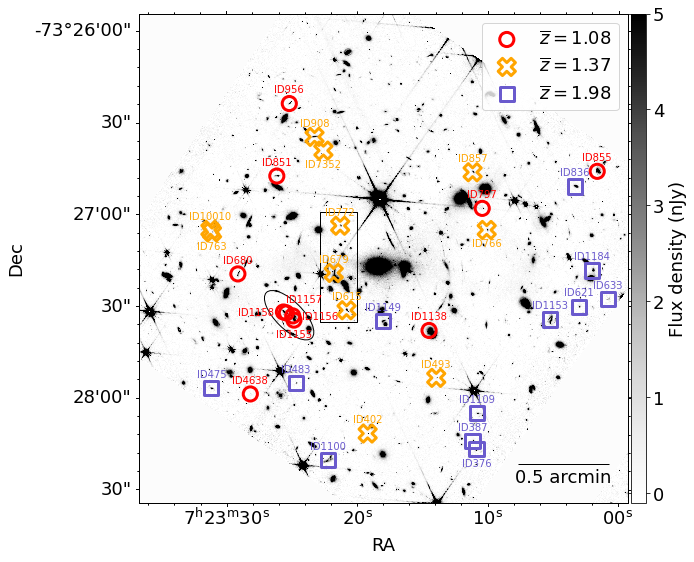}
    \includegraphics
    [width=0.49\textwidth]
    {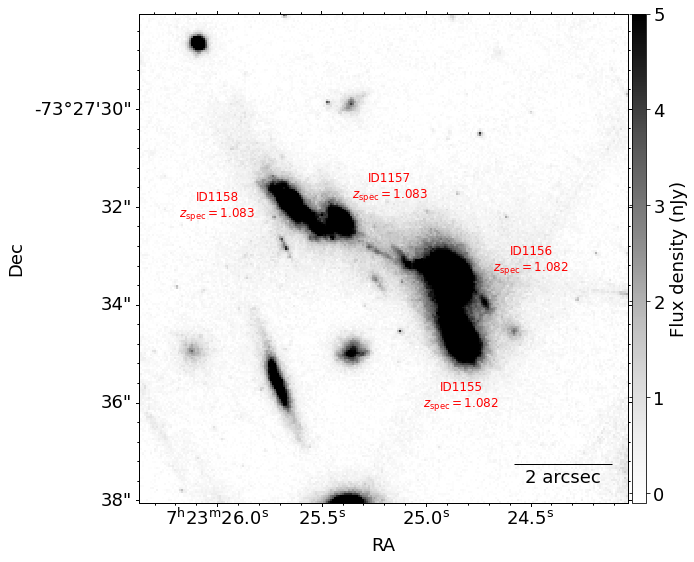}
    \caption{
    {\it Left}: Spatial distribution of the three overdensities at $\overline{z}=1.08$ (red circles), $\overline{z}=1.37$ (orange crosses), and $\overline{z}=1.98$ (slate-blue squares). The black rectangle indicates the three images of the Sparkler galaxy, and the black ellipse indicates the four closely-interacting galaxies in the $\overline{z}=1.08$ overdensity. {\it Right}: Close-up view of the four closely-interacting galaxies in the $\overline{z}=1.08$ overdensity. The four sources have MUSE spectroscopic redshifts within $z=1.082-1.083$. Source ID~1155 has two MUSE components (IDs~1155.1 and 1155.2), including one with a consistent grism redshift measurement (ID~1155.2). Both panels show the $0\farcs04$ NIRCam F150W imaging.  
    }
    \label{fig:spatial_dist}
\end{figure*}

\begin{figure*}
    \centering
    \fbox{\includegraphics[width=1\textwidth]{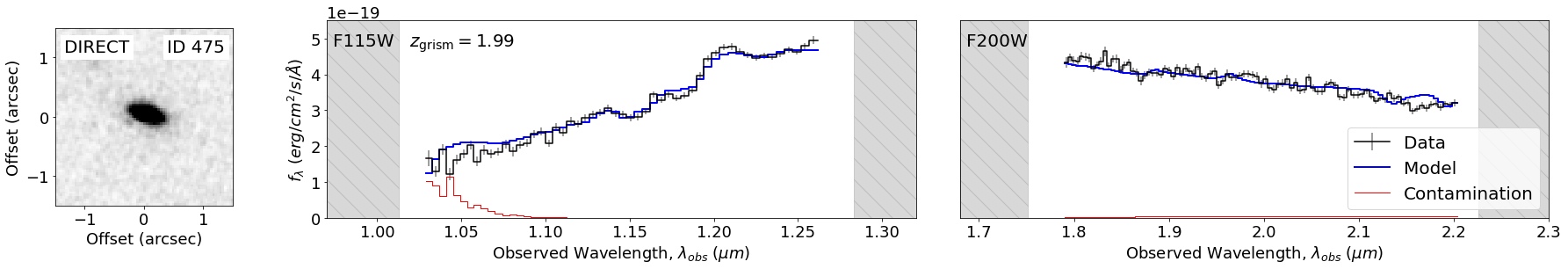}}
    \fbox{\includegraphics[width=1\textwidth]{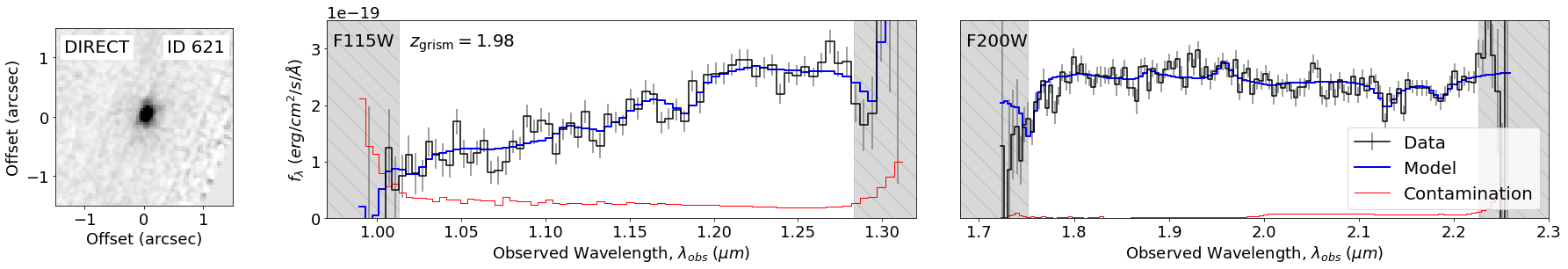}}
    \fbox{\includegraphics[width=1\textwidth]{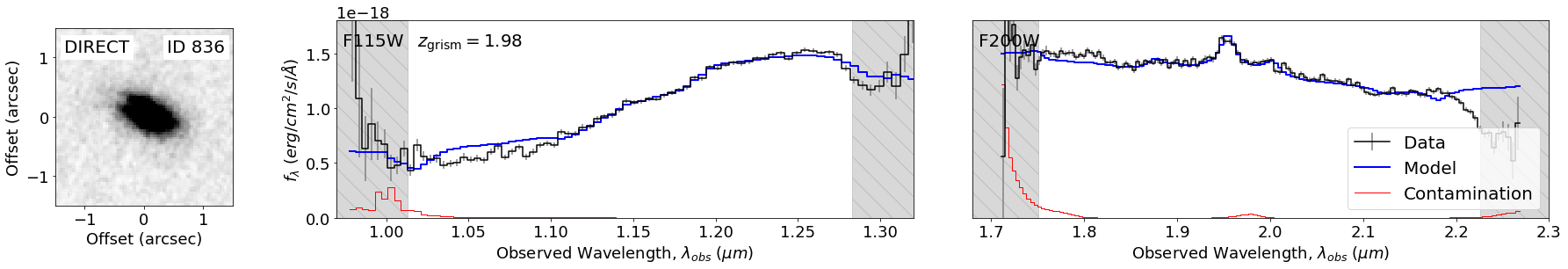}}
    \caption{Direct image stamps and 1D grism spectra of the three sources in the $\overline{z}=1.98$ overdensity with significant spectral breaks in their F115W grism data, suggestive of old, quiescent stellar populations. In all panels, the contamination-subtracted F115W and F200W data are shown with black solid lines, while the best-fitting models and contamination estimates are shown in blue and red, respectively. Source IDs~475 (top), 621 (middle), and 836 (bottom) have grism redshifts of $z_{\rm grism}=1.99$, $1.98$, and $1.98$, respectively.}
    \label{fig:grism_spectra_breaks}
\end{figure*}

\item {\bf $\overline{z}=1.37$ overdensity.} In the second overdensity ($\overline{z}=1.37$), we identify nine sources within $z=1.36-1.39$, including one galaxy imaged three times by the foreground lensing cluster (artificially increasing the number of counts to 11 in Figure~\ref{fig:redshift_histogram}). The multiple images and redshift of this galaxy were first reported in \citet{Golubchik2022}. We here adopt the redshift of $z = 1.378 \pm 0.001$ obtained from the MUSE [OII]$\lambda3727$ line \citep{Caminha2022, Golubchik2022, Mahler2022arXiv}. In \citet{Mowla2022}, we further report on the discovery and analysis of twelve compact sources associated with the main body of the galaxy (refered to as the ``Sparkler''). In that work, we analyse the color of the compact sources as well as their spectral energy distributions and find that five of the twelve sources are consistent with being evolved globular clusters comprised of single stellar populations formed at $z>9$, i.e., within in the first 500 Myr of cosmic history. The three images of the Sparkler are included in our final catalogue (IDs~615, 679, and 772, corresponding to images 3, 2, and 1 of the Sparkler in the lensing models, respectively). The redshift of the other eight sources comprising the overdensity are based solely on our NIRISS grism analysis. Among those eight sources, one (ID~908) is comprised of three grism components. Two of the components have measured redshifts of $z=1.370$ and $1.372$ (IDs 908.1 and 908.3, respectively), while the third (ID 908.2) has a measured redshift of $z=1.342$. The lower redshift of this component could be due to a relative offset between the direct image centroid and the line-emitting region, contamination from another source, or a real offset indicating that this component is a foreground object. In our final, merged catalogue (c.f.~Table~\ref{tab:catalog}), we report the mean redshift of the three components as the redshift of the object (i.e., $z=1.362$). The spatial distribution of the overdensity is presented in the left panel of Figure~\ref{fig:spatial_dist} (orange crosses). On this Figure, the three images of the Sparkler galaxy are indicated by a black rectangle. Note that we have tentatively identified a number of additional sources in the field that are potentially associated with this $\overline{z}=1.37$ overdensity. Further analysis of these sources is necessary to verify the nature of the features seen in their grism spectra, and this is beyond the scope of the current paper. A more detailed study of this overdensity and its members is deferred to a later analysis.

\begin{figure*}
    \centering
    \includegraphics
    [width=1\textwidth]
    {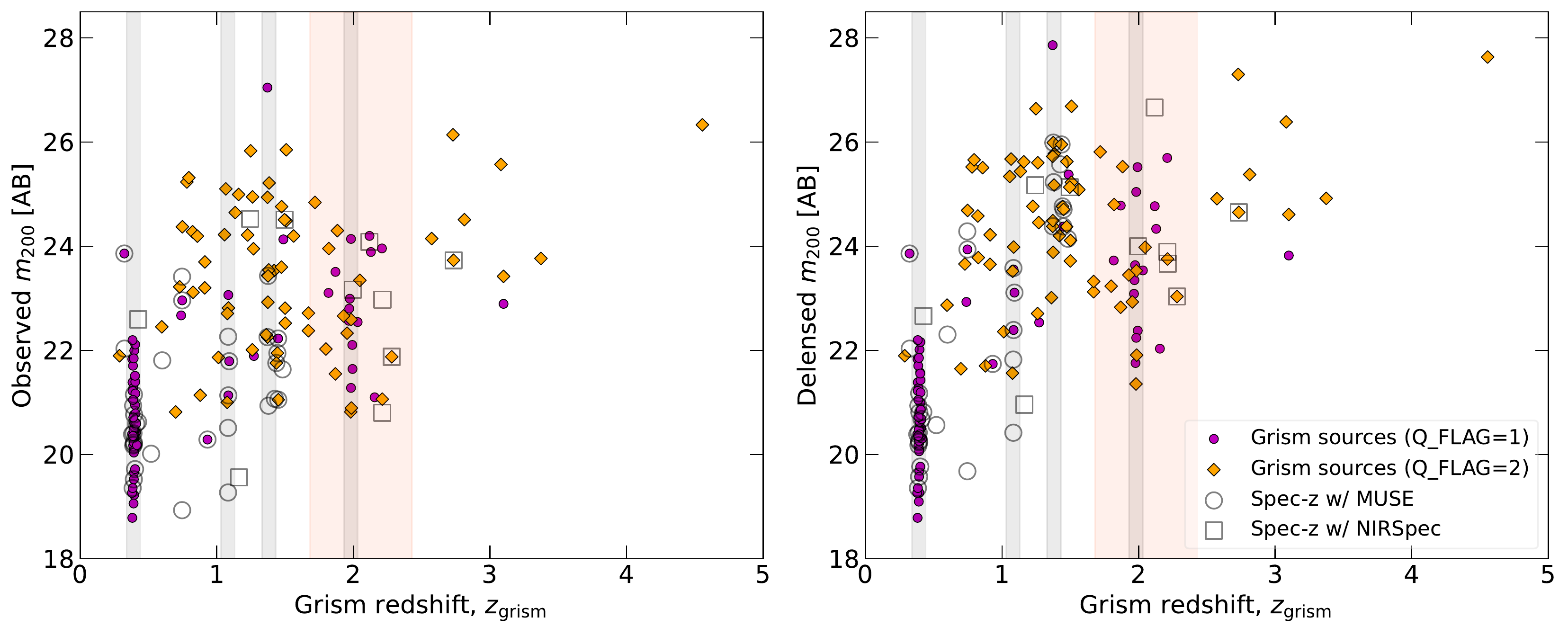}
    \caption{
    Magnitude-redshift diagram for sources in our final spectroscopic catalogue, before lensing correction ({\it left}) and after lensing correction ({\it right}). The apparent magnitudes are measured in the F200W band. The magenta circles and orange diamonds represent grism sources of quality flags 1 and 2, respectively. The error bars on the data points are smaller than the symbols except for only one grism source and are therefore not shown in this Figure. Sources with MUSE and NIRSpec spectroscopic redshifts are also indicated with open circles and open squares, respectively. Sources with $z_{\rm spec}>5$ are not shown in the diagram. The cluster redshift ($z_{\rm cl}=0.39$) and the three additional redshift overdensities at $z\sim1.1$, $1.4$, and $2.0$ are highlighted by the vertical grey shaded areas. The redshift range where multiple, typically strong emission lines can be observed with the NIRISS data of the field ($1.68<z<2.43$) is shown by a light-red shaded area.  
    }
    \label{fig:mag_redshift}
\end{figure*}

\item {\bf $\overline{z}=1.98$ overdensity.} As mentioned in Sec.~\ref{sec:line_coverage} (and Fig.~\ref{fig:line_coverage}), the redshift range $z=1.68-2.43$ (highlighted by the light-red shaded area on Figure~\ref{fig:redshift_histogram}) is the only range where two typically strong emission-lines are visible at the same time in the F115W and F200W filters ([OII]$\lambda3727$ and H$\alpha$). In that range, this effect might potentially bias the number of sources for which we obtain a secure NIRISS redshift. However, the $\overline{z}=1.98$ overdensity spans a significantly smaller redshift range, indicating that it is a real overdensity and not an instrumental bias. We identify twelve sources within $z=1.96-2.00$, including one based on the NIRSpec data, and eleven based solely on our NIRISS grism analysis. Remarkably, three of the sources (IDs~475, 621, 836) show significant spectral breaks in their F115W grism data, suggestive of old, quiescent stellar populations. We show the 1D grism spectra of these three sources in Figure~\ref{fig:grism_spectra_breaks}. The colors of these three galaxies, as well as an additional source of the overdensity, are also consistent  with $z=2.0$ quiescent galaxies (see Sec.~\ref{sec:colors}). 
Among the eleven grism sources of the overdensity, two have multiple components identified in our work. One source (ID~376) possesses three components within $\sigma_z=\pm 0.008$ of $\overline{z} = 1.966$, while one source (ID~1153) possesses two components within $\pm 0.001$ of $1.985$. Interestingly, the twelve sources of this overdensity seem to be mostly located in the South part of the field, as is shown by the slate-blue squares in Figure~\ref{fig:spatial_dist}.

\item {\textbf{Spatial distribution of the overdensities.}} 
{In the left panel of Figure~\ref{fig:spatial_dist}, we present the spatial distribution of the sources in each of the newly-identified overdensities. Red circles indicate sources of the $\overline{z}=1.08$ overdensity (selected within $z_{\rm spec, grism} = 1.05-1.09$), orange crosses indicate sources of the $\overline{z}=1.37$ overdensity (selected within $z_{\rm spec, grism} = 1.36-1.39$), and slate-blue squares indicate sources of the $\overline{z}=1.98$ overdensity (selected within $z_{\rm spec, grism} = 1.96-2.00$). On the Figure, the black rectangle indicates the three images of the Sparkler galaxy, and the black ellipse shows the position of the four closely-interacting sources of the $\overline{z}=1.08$ overdensity. We show a close-up view of these four sources in the right panel of the Figure.
Interestingly, the identification of the strongly-lensed, multiply-imaged Sparkler galaxy in the $\overline{z}=1.37$ overdensity, as well as the roughly even distribution of the other $\overline{z}=1.37$ sources around the $z_{\rm cl}=0.39$ brightest cluster galaxy (BCG), suggest that this overdensity may be relatively well aligned with the foreground $z_{\rm cl}=0.39$ cluster, potentially increasing the chances of strong lensing magnification. This makes this overdensity a potentially good target for spatially resolved studies of its members. While we do not identify any strongly-lensed, multiply-imaged system in the $\overline{z}=1.08$ overdensity, the spatial distribution of its members around the $z_{\rm cl}=0.39$ BCG may also suggest a relatively good alignment of this overdensity with the foreground cluster. On the other hand, the sources of the $\overline{z}=1.98$ overdensity seem to be mostly located in the South-West region of the field. The chances of strong-lensing magnification for the sources of this overdensity may therefore be less than that of the other two.}

\end{itemize}

In addition to the well-known cluster at $z_{\rm cl}=0.39$, our work identifies three galaxy overdensities, at $\overline{z}=1.08$, $1.37$, and $1.98$. These are easily discernible in Fig.~\ref{fig:redshift_histogram} and are also apparent in the left panel of Fig.~\ref{fig:mag_redshift}, which shows the observed (i.e., not corrected for lensing magnification) F200W magnitude vs.~redshift of the sources in our final catalogue.  On the other hand, the right panel of Fig~\ref{fig:mag_redshift} shows the magnitude-redshift diagram for F200W magnitudes that have been corrected for lensing magnification using the $\mu$ values from the \citet{Caminha2022} model. This Figure illustrates how deep in the galaxy population our catalogue probes. As mentioned earlier, the Sparkler galaxy, which is magnified by perhaps as much as 100$\times$ \citep{Mowla2022}, is part of the $\overline{z}=1.37$ overdensity seen in Figs.~\ref{fig:redshift_histogram} and \ref{fig:mag_redshift}. Given that many of the galaxies in the three redshift overdensities are significantly magnified by the gravitational potential of the foreground $z_{\rm cl}=0.39$ cluster, these structures can be interesting follow-up targets for studies of galaxies in cluster or group environments at higher redshifts. 

Overall, our work increases the number of secure spectroscopic redshifts in the field of SMACS~J0723.3$-$7327 by a factor of three compared to previous studies, and by more than an order of magnitude between our new {\it JWST}/NIRISS redshifts and previously published {\it JWST}/NIRSpec data.

\subsection{Colour distributions}\label{sec:colors}

\begin{figure}
	\includegraphics[width=\columnwidth]{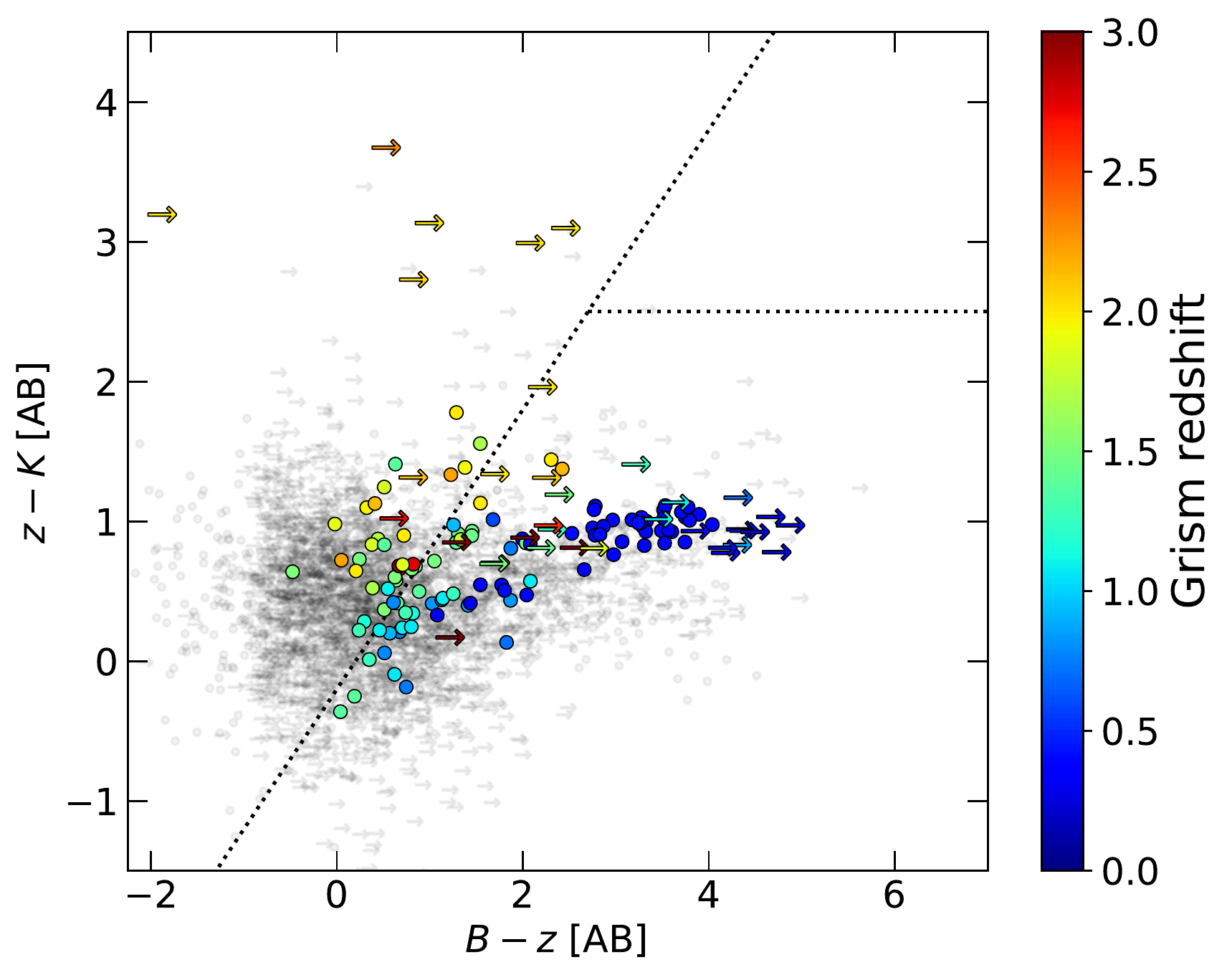}
    \caption{
    The $B\! z\! K$ diagram (measured directly from F435W, F090W, and F200W fluxes) for sources in our photometric catalogue (grey points) and in our final grism redshift sample (coloured according to $z_{\rm grism}$). Only sources with $\rm S/N > 3$ in the different filters are shown in this diagram. Sources not detected in the F435W image are shown with right arrows. The dotted lines show the regions from \citet{daddi_new_2004} used to identify $z<1.4$ galaxies (lower right corner), $z>1.4$ star-forming galaxies (left-hand side of the diagonal line), and $z>1.4$ quiescent galaxies (upper right wedge). 
    }
    \label{fig:BzK}
\end{figure}
\begin{figure}
	\includegraphics[width=\columnwidth]{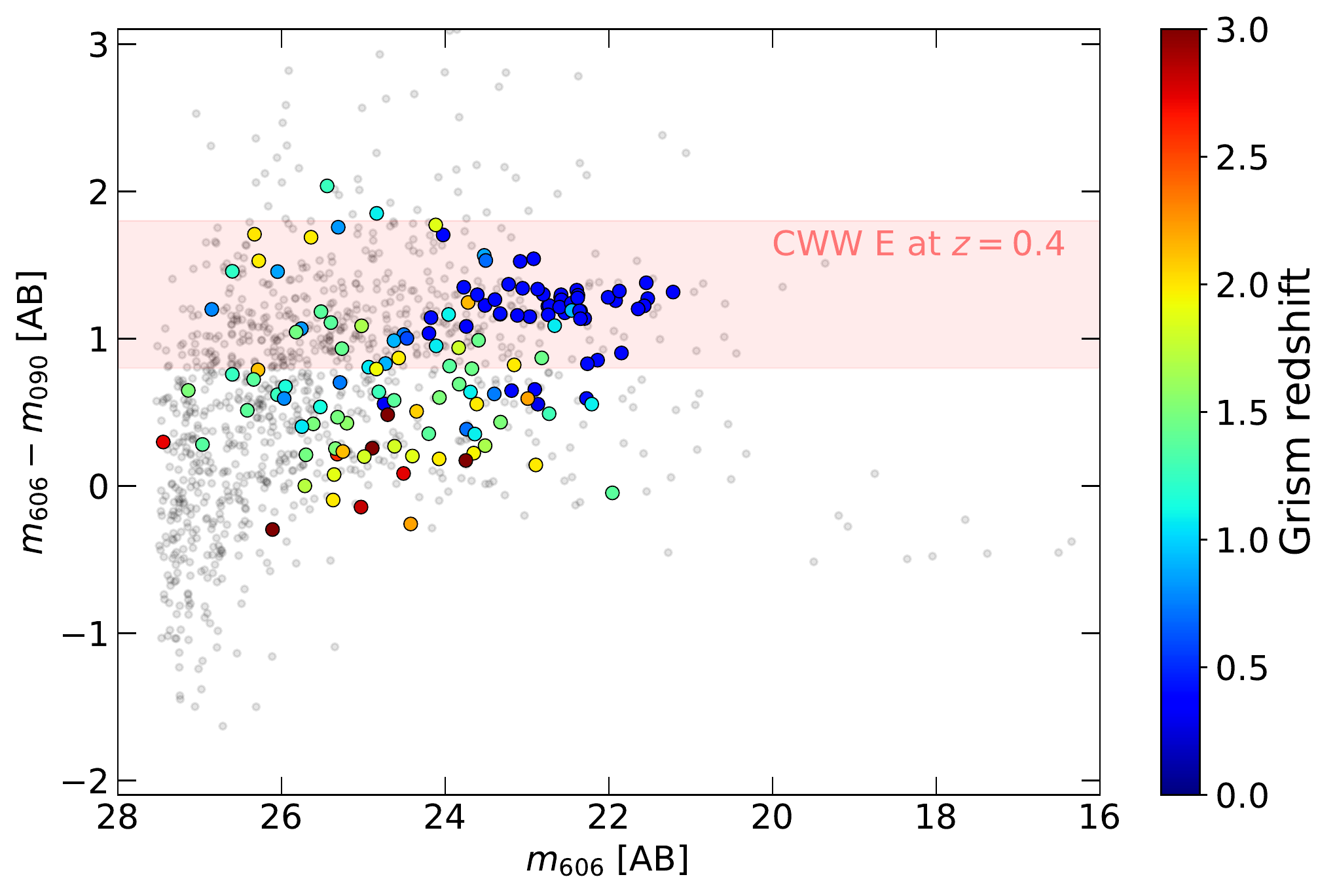}
	\includegraphics[width=\columnwidth]{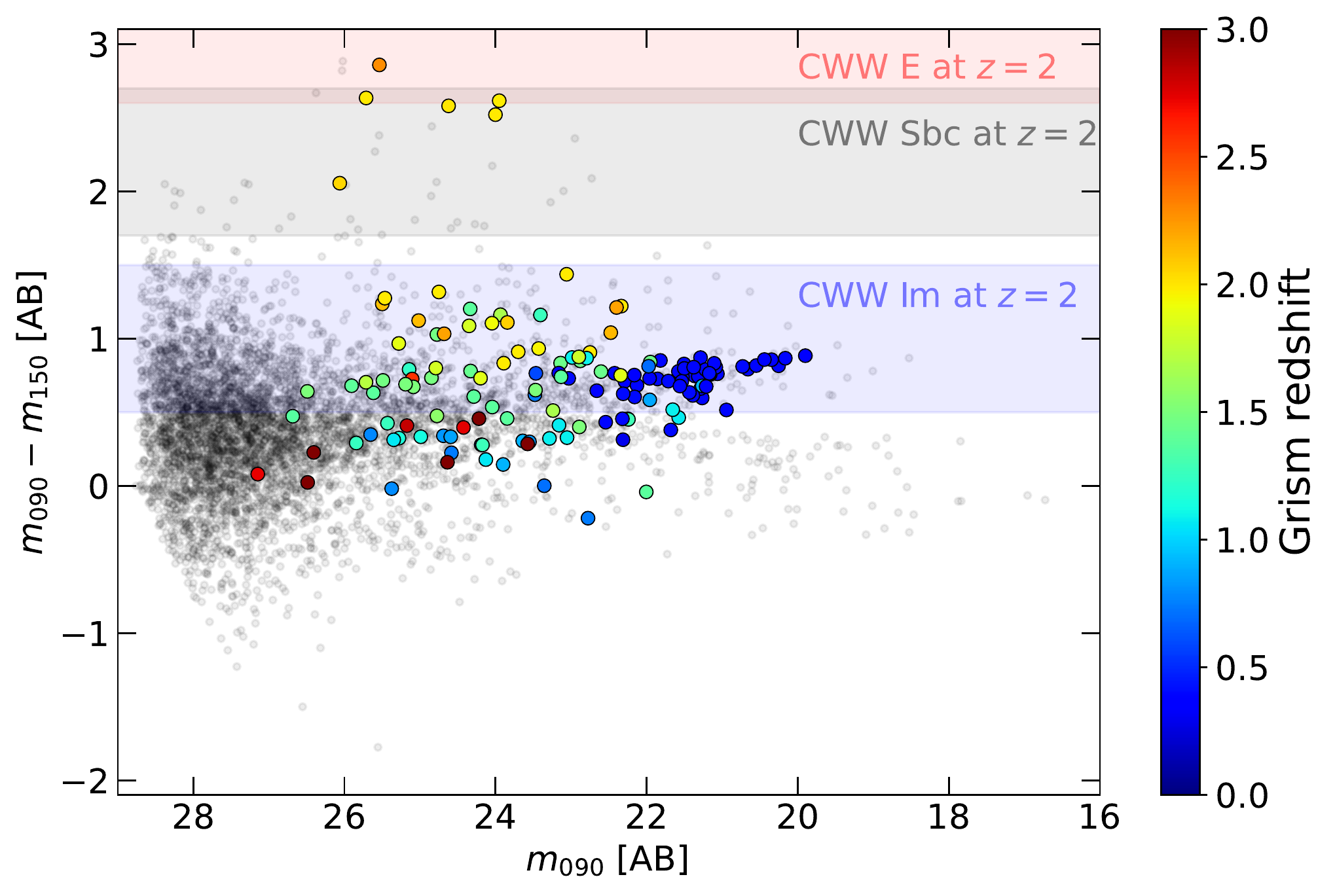}
    \caption{
    The color-magnitude diagrams (CMDs) for sources in our photometric catalogue (grey points) and in our final grism redshift sample (coloured according to $z_{\rm grism}$). Only sources with $\rm S/N > 3$ in the different filters are shown in the CMDs.
    {\it Top panel}: the $m_{606}-m_{090}$ vs.~$m_{606}$ CMD. These colours roughly bracket the D4000 break at the redshift of the $z_{\rm cl}=0.39$ cluster. The red shaded area shows the expected colours of the cluster ellipticals, assuming magnitude uncertainties of $\Delta m=\pm0.25$, calculated from the \citet{coleman_colors_1980} spectral template of elliptical galaxies.
    {\it Bottom panel}: the $m_{090}-m_{150}$ vs.~$m_{090}$ CMD. These colours roughly bracket the D4000 break at the redshift of the $\overline{z}=1.98$ overdensity. In this panel, the red, grey, and blue shaded regions show the expected colours of the \citet{coleman_colors_1980} spectral templates of E, Sbc, and Im galaxies at $z=2$, respectively, again assuming magnitude uncertainties of $\Delta m=\pm0.25$.
    }
    \label{fig:CMD}
\end{figure}

Figure~\ref{fig:BzK} shows the observed-frame $BzK$ diagram \citep{daddi_new_2004} of all sources in our final catalogue (colour-coded by their grism redshift) as well as of sources from our parent photometric dataset (grey points). We directly use the observed {\it HST}/ACS F435W, and {\it JWST}/NIRCam F090W, and F200W fluxes that probe the $B$, $z$, and $K$ pass-bands, respectively, and only show sources that have a ${\rm S/N}>3$ in the different filters. Sources that are not detected in the F435W filter (i.e., $B$ pass-band) are shown as right arrows in the Figure. We also show as dotted lines the separation between $z>1.4$ star-forming (left-hand side of the diagonal line), $z>1.4$ quiescent (upper-right wedge), and $z<1.4$ (lower-right corner) galaxies \citep{daddi_new_2004}. As expected, galaxies identified at the cluster redshift populate the lower-right corner of the diagram, and galaxies with $z_{\rm grism}>1.4$ are predominantly located on the left-hand side of the diagonal line. A number of $z_{\rm grism} \geq 2.0$ sources with red, $z - K > 2.5$~mag (AB), colours are not detected in the F435W observations and it cannot be determined whether their colours are consistent with that of $z>1.4$ quiescent galaxies based on the $BzK$ diagram alone.

In Figure \ref{fig:CMD}, we show the colour-magnitude diagrams (CMDs) of the sources in our final catalogue (colour-coded by their grism redshift) and of that of our parent photometric dataset (grey points). In the top panel of the Figure we show the $m_{606}-m_{090}$ vs.~$m_{606}$ diagram (roughly probing the D4000 break at the redshift of the cluster), while we show the $m_{090}-m_{150}$ vs.~$m_{090}$ diagram (roughly probing the D4000 break at the redshift of the $\overline{z}=1.98$ overdensity) in the bottom panel. Again, we only show sources that are detected with ${\rm S/N}>3$ in the different filters. The horizontal shaded areas show the expected colours of elliptical, spiral, and irregular galaxies derived from the Elliptical, Sbc, and Im template spectra of \citet{coleman_colors_1980} (CWW) at the redshifts indicated in the Figure.

In the top panel of Figure~\ref{fig:CMD}, the $z_{\rm cl}=0.39$ SMACS~J0723.3$-$7327 cluster quiescent galaxies (blue circles) are seen forming a cluster red sequence at $m_{606}-m_{090}\sim 1.5$~mag (AB), as expected from the CWW models. These objects are also seen as a tight sequence in the $m_{090}-m_{150}$ vs.~$m_{090}$ colour-magnitude diagram, shown in the bottom panel of Fig.~\ref{fig:CMD}. However, this sequence is not the classic cluster color-magnitude `red-sequence' since the D4000 break is at shorter wavelengths than those probed by $m_{090}-m_{150}$ colours, but is due to the uniform and relatively flat colour of cluster galaxies above the D4000 break. Based on our template set of old, quiescent galaxy spectra, we interpret the tilt in this $z\sim0.4$ color-magnitude sequence as due to the metallicity and age dependence on galaxy stellar mass since the $m_{090}-m_{150}$ colour (rest-frame $\lambda_{\rm rest} = 0.6 - 1.1$~\micron~at the redshift of the cluster) is sensitive to the strength of the TiO bands present within that wavelength range. 
The Figure illustrates that even relatively shallow NIRISS observations can secure large samples of absorption-line redshifts for intermediate-redshift galaxies.

The bottom panel of Fig.~\ref{fig:CMD} also reveals six $z\sim2$ sources with red, $m_{090}-m_{150}>2$~mag (AB), colours, consistent with the colours of quiescent galaxies at these redshifts. Among them are the three sources with significant D4000 breaks in their NIRISS spectra (see Fig.~\ref{fig:grism_spectra_breaks}), which are associated with the $\overline{z}=1.98$ redshift overdensity. We interpret them as group or (proto)cluster quiescent galaxies at this redshift.
These six red, $z\sim2$ objects also stand out as $z-K>2.5$~mag (AB) objects in the $BzK$ diagram shown in Fig.~\ref{fig:BzK}. As already mentioned, none of the six are detected in the $B$ pass-band (observed F435W), but their $z-K$ colours and $B-z$ limits are not inconsistent with the colours of $z\sim2$ quiescent galaxies. This is not inconsistent with our observation that at least three of these objects are high-redshift quenched galaxies associated with the $\overline{z}=1.98$ redshift overdensity.

In addition to the red, quiescent sources with prominent D4000 breaks in their NIRISS spectra, the $\overline{z}=1.98$ overdensity also contains a number of sources consistent with the colours of $z\sim2$ star-forming galaxies (consistent with the $z=2$ CWW Im spectral template in Fig.~\ref{fig:CMD}, shown as the light-blue shaded area) and identified through the presence of [OII]3727\AA\ and H$\alpha$ emission lines in their spectra. This $\overline{z}=1.98$ galaxy overdensity, containing a significant number of both red quiescent and blue star-forming galaxies offers the opportunity to study in great detail the properties of galaxies in a dense environment at cosmic noon.

\section{Caveats}
\label{sec:caveats}

As mentioned and demonstrated in the previous sections, we have taken a conservative approach to obtain secure spectroscopic redshifts for a fairly large subset of sources that can be of use to the community, rather than a complete catalogue for the full sample. Objects with redshifts that we consider to be uncertain are thus excluded from our final catalogue. While our final spectroscopic catalogue is deemed secure, we here list the different caveats associated to our underlying data and analysis.

\begin{itemize}

    \item {\bf Photometry.} In this work, we use a pre-flight reference file for which photometric zero-points may be incorrect \citep[see, e.g.,][]{Boyer2022}. As mentioned, we therefore use {\tt EAzY} to re-derive proper photometric zero-points consistent with the photometric redshift fitting of our source catalogue as we describe in \citet{Mowla2022}, and we also add a systematic error of $3\%$ of the flux to the photometric errors to derive our photometric redshifts.
    To test the quality of the photometry, we use the spectroscopic redshifts as well as the secure quality 1 grism redshifts to independently re-derive new zero-point corrections using {\tt Phosphoros}. In all {\it JWST}/NIRCam bands, the new corrections are less than $5\%$ of the fluxes. This is also true for the {\it HST}/ACS bands, except for the F435W where the correction reaches $\sim +40\%$. However, this large correction factor can be explained by the shallow depth of the {\it HST}/ACS F435W observations and the large errors on the F435W fluxes which do not allow a proper measurement of the zero-point correction for this filter. As an additional test, we derive new photometric redshifts using the {\tt Phosphoros} zero-point corrections, and we notice only small differences in the results, indicating that the large correction factor in the F435W band does not have a strong impact and that the errors on the flux measurements dominate. {Quantitatively, the variation in the scatter and outlier fraction of our photometric redshifts when using the {\tt Phosphoros} zero-point corrections is below $1\%$. Using the {\tt Phosphoros} zero-point corrections instead of the {\tt EAzY} zero-points would therefore not change our results.}

    \item {\bf Completeness and purity.} As shown in the comparison of independently-measured grism and spectroscopic redshifts in Section~\ref{sec:redshift_fidelity}, our final spectroscopic catalogue represents a high-quality sample of redshifts, for both NIRISS grism redshifts of quality 1 (multiple features) and of quality 2 (single features identified with our photometric redshifts). Our redshift fidelity assessment suggests that very few, if any, of the redshifts are catastrophically wrong. Our catalogue is therefore deemed very secure, but at a potential high-cost against completeness, in terms of e.g., magnitude, colour, redshift, etc. Note that the lack of the F150W filter in the NIRISS WFSS observations of the SMACS~J0723.3$-$7327 field (see Sec.~\ref{sec:line_coverage} and Fig.~\ref{fig:line_coverage}), as well as the spectral contamination due to the high density of sources, additionally conspire against completeness in the field.

    \item {\bf Redshift precision and accuracy.} Due to the low resolution of the two orthogonal grisms ($R\sim150$) and the morphological broadening of emission and absorption features for extended objects\footnote{While we are developing methods to properly model and fit the grism data of such large objects (see, e.g., Sec.~3.3 in \citealt{Mowla2022}, and Estrada-Carpenter et al.~in prep.), those are currently not included in the present catalogue.}, the NIRISS grism redshifts of this work are of lower precision than the NIRSpec ($R\sim 1000$) and MUSE ($R\sim 3000$) redshifts of the field.
    Additionally, as shown in Section~\ref{subsection:field-variation}, {we observe a slight redshift trend of the order of $|{\Delta z_{\rm grism}}| \leq 0.01$ with respect to the North-South direction for galaxies around the cluster redshift. While this trend is not seen when comparing our grism redshifts to the spectroscopic redshifts of the field covering the central part of the NIRISS field-of-view, we cannot completely rule out that} the current reduction of the NIRISS data may be affected by slight field-dependent calibration issues (e.g., slight wavelength-solution systematics). While further analysis is necessary to better understand the origin of this slight field-variation, we caution the reader that we conservatively assume the current accuracy of our grism redshift measurements to be no better than $\delta z \sim 0.01$. {In Table~\ref{tab:catalog} \&~\ref{tab:catalog2}  the reported redshift uncertainties are the $1\sigma$ errors from {\tt Grizli}'s redshift PDFs; i.e., the redshift precision from the fitting.}

    \item {\bf Cluster galaxies.} In this work, the NIRISS grism redshifts of the bright cluster ellipticals are obtained based on the identification and fitting of the multiple, strong (TiO) absorption features present in the spectrum of old, quiescent galaxies around $\lambda_{\rm rest} \sim 0.8~\micron$ (see Fig.~\ref{fig:grism_spectra_clg}, and \citealp{GrayCorballyBook}). Given that a number of absorption bands are present in this spectral region, it is possible that a number of foreground or background objects have been misidentified for $z_{\rm cl}=0.39$ cluster galaxies. However, the comparison of our grism redshifts to the spectroscopic redshifts in the field do not reveal any such mis-identification (see Fig.~\ref{fig:zgrism_vs_zspec}). This suggests that our final spectroscopic catalogue is little to not affected by misidentified cluster galaxies.
    
    \item {\bf Multiply-imaged sources.} Because of the gravitational-lensing effect of the SMACS~J0723.3$-$7327 cluster, some sources in our final spectroscopic catalogue may correspond to the same, multiply-imaged objects. This may potentially affect the number of individual galaxies shown in our redshift distribution and in the three newly-identified redshift overdensities. In the $\overline{z} = 1.37$ overdensity, one source (the Sparkler) is imaged three times. We compare the observed colours of the other sources in the three overdensities, and visually inspect their direct image stamps, but we do not find evidence for additional multiply-imaged sources. Also, we do not find any match between our new NIRISS grism sources and the list of multiple-image systems of \citet{Caminha2022} and \citet{Mahler2022arXiv}. The multiple-image systems with a match in our final catalogue are spectroscopic sources with already known MUSE or NIRSpec redshifts (including the Sparkler), except system 4 in the \citet{Caminha2022} model for which we identify a new, previously unpublished, NIRSpec redshift (see Appendix~\ref{sec:methods-nirspecredshifts}). This comparison suggests that the new NIRISS grism redshifts identified in our work are likely comprised of a very small number, if any, of multiply-imaged systems. In particular, this should not affect our conclusions regarding the three newly-identified redshift overdensities in the field.

    \item {\bf Mismatched sources.} In our final spectroscopic catalogue, we merge together grism sources falling within the same segmentation region of a single photometric object (see Sec.~\ref{sec:methods-grism-photoz}). In this step, we verify {\it a posteriori} that the grism redshift measurements of the different grism components are consistent with one another. In all cases, we find that the different grism components of a single photometric source agree within $\pm 0.01\times(1+z)$. This suggests that only a very small number of sources, if any, may potentially be affected by a mis-match between our photometric and spectrocopic catalogues.
        
\end{itemize}

\section{Conclusions}\label{sec:conclusion}


In this paper, we present a catalogue of spectroscopic redshifts in the field of SMACS~J0723.3$-$7327. We primarily use the {\it JWST}/NIRISS slitless grism spectroscopy of the Early Release Observations of the field, supported with {\it JWST}/NIRCam and {\it HST}/ACS photometry. And we complement our grism spectroscopic redshifts with {\it JWST}/NIRSpec and VLT/MUSE redshifts published in the literature. Throughout this work, we take a conservative approach to obtain secure NIRISS grism redshifts. To do so, we first develop a multi-step methodology to review and assess the quality of the NIRISS grism data and of our redshift fitting procedure. This methodology, which includes the visual inspection of the data products and of each spectral fitting by multiple reviewers, is primarily based on ({\it i}) the unambiguous identification of multiple spectral features in the NIRISS grism data (quality 1) or ({\it ii}) the unambiguous identification of a single spectral feature whose nature is securely determined based on two independent photometric redshift estimates (namely, using {\tt Phosphoros} and {\tt EAzY}; quality 2).
Based on this careful methodology, we obtain the following results:

\begin{itemize}
\item[1.] We obtain a spectroscopic catalogue in the field of SMACS~J0723.3$-$7327 consisting of a total of 190 sources with secure spectroscopic redshifts, including 156 NIRISS grism redshifts, 123 of which are for sources whose redshifts were previously unknown. This increases the number of secure spectroscopic redshifts in this field by a factor of three compared to previous studies, and by more than an order of magnitude between our new {\it JWST}/NIRISS redshifts and previously published {\it JWST}/NIRSpec data.

\item[2.] The redshift distribution of our final spectroscopic catalogue recovers the redshift peak at the redshift of the SMACS~J0723.3$-$7327 cluster ($z_{\rm cl} = 0.39$). In particular, the new NIRISS grism redshifts that we obtain in this work more than double the number of spectroscopic members of the cluster, which we identify based on the strong TiO absorption features ($\lambda_{\rm rest} \sim 0.8~\micron$) seen in their F115W grism data. It also increases the number of background sources behind SMACS~J0723.3$-$7327 from 31 to 117, including the five new $z>0.5$ NIRSpec redshifts we obtain in our work.

\item[3.] We identify three galaxy overdensities behind SMACS~J0723.3$-$7327 thanks to our significant increase in the number of  secure spectroscopic redshifts in the field. These newly-discovered overdensities, until now missed by the existing {\it JWST}/NIRSpec and VLT/MUSE data, are at $\overline{z} = 1.08$, $1.37$, and $1.98$, with $\gtrsim 10$ secure spectroscopic members each. In particular, we identify twelve sources in the $\overline{z} = 1.98$ overdensity, eleven of which are based solely on our NIRISS grism analysis. Three of the sources of this overdensity also show significant spectral breaks in their F115W grism data, consistent with $z=2.0$ quiescent galaxies.

\item[4.] We compare {\it a posteriori} our NIRISS grism redshifts to the spectroscopic redshifts of the field. We find an excellent agreement between grism and spectroscopic redshifts, for both quality 1 and quality 2 grism sources, with all sources but one showing deviations of less than $\pm0.01\times(1+z_{\rm spec})$ with marginal systematic bias. Using the cluster galaxy grism redshifts, we however find a slight systematic trend no larger than $\pm0.01$ in redshift with respect to the North-South direction of the field. {While this could be a real trend}, we conservatively assume the current accuracy of our grism redshift measurements to be no better than $\delta z \sim 0.01$.

\item[5.] We find that the colour distribution of the cluster galaxies identified in our NIRISS analysis is consistent with the expected colours of elliptical galaxies at the redshift of the cluster. Additionally, we find that the colours of six of our sources at $z\geq2$ are consistent with that of $z=2.0$ old, quiescent galaxies, including the three galaxies of the $\overline{z} = 1.98$ overdensity with strong spectral breaks in their grism data.

\end{itemize}

Altogether, our work demonstrates the power of {\it JWST}, and particularly of the Wide Field Slitless Spectroscopy mode of the NIRISS instrument to identify and securely determine the redshift of large numbers of extragalactic sources, even in relatively shallow NIRISS grism exposures that were further impacted by the lack of the middle F150W filter in these specific observations. Our spectroscopic catalogue\footnote{Which is made publicly available at \url{https://niriss.github.io/smacs0723}.} offers a new, significant resource of carefully-vetted spectroscopic redshifts in ``Webb's First Deep Field'', SMACS~J0723.3$-$7327. It also offers a basis for further studies of the properties of the new cluster galaxies identified in our work, as well as of the background sources, in particular those of the three new overdensities identified in our work.

\section*{Acknowledgements}


{We thank the referee for their useful comments and suggestions that improved the clarity of the paper.} This research was enabled by grant 18JWST-GTO1 from the Canadian Space Agency, funding from the Natural Sciences and Engineering Research Council of Canada, and scholarships to YA from from the Japan Society for the Promotion of Science. {This research used the Canadian Advanced Network For Astronomy Research (CANFAR) operated in partnership by the Canadian Astronomy Data Centre and The Digital Research Alliance of Canada with support from the National Research Council of Canada the Canadian Space Agency, CANARIE and the Canadian Foundation for Innovation.}

{\it Software}: {\tt ASTROPY} \citep{Astropy2018}, {\tt MATPLOTLIB} \citep{Hunter2007}, {\tt NUMPY} \citep{Harris2020}, {\tt SCIPY} \citep{Virtanen2020}.

\section*{Data Availability}


Grism redshifts and associated photometry measurements made as part of this work are included as online material to this paper and are also available from the CANUCS project website ({\url{https://niriss.github.io/smacs0723}}).  The underlying {\it JWST} and {\it HST} data are publicly available from MAST.



\bibliographystyle{mnras}
\bibliography{references}

\begin{thebibliography}{}
\makeatletter
\relax
\def\mn@urlcharsother{\let\do\@makeother \do\$\do\&\do\#\do\^\do\_\do\%\do\~}
\def\mn@doi{\begingroup\mn@urlcharsother \@ifnextchar [ {\mn@doi@}
  {\mn@doi@[]}}
\def\mn@doi@[#1]#2{\def\@tempa{#1}\ifx\@tempa\@empty \href
  {http://dx.doi.org/#2} {doi:#2}\else \href {http://dx.doi.org/#2} {#1}\fi
  \endgroup}
\def\mn@eprint#1#2{\mn@eprint@#1:#2::\@nil}
\def\mn@eprint@arXiv#1{\href {http://arxiv.org/abs/#1} {{\tt arXiv:#1}}}
\def\mn@eprint@dblp#1{\href {http://dblp.uni-trier.de/rec/bibtex/#1.xml}
  {dblp:#1}}
\def\mn@eprint@#1:#2:#3:#4\@nil{\def\@tempa {#1}\def\@tempb {#2}\def\@tempc
  {#3}\ifx \@tempc \@empty \let \@tempc \@tempb \let \@tempb \@tempa \fi \ifx
  \@tempb \@empty \def\@tempb {arXiv}\fi \@ifundefined
  {mn@eprint@\@tempb}{\@tempb:\@tempc}{\expandafter \expandafter \csname
  mn@eprint@\@tempb\endcsname \expandafter{\@tempc}}}

\bibitem[\protect\citeauthoryear{{Astropy Collaboration} et~al.,}{{Astropy
  Collaboration} et~al.}{2018}]{Astropy2018}
{Astropy Collaboration} et~al., 2018, \mn@doi [\aj] {10.3847/1538-3881/aabc4f},
  \href {https://ui.adsabs.harvard.edu/abs/2018AJ....156..123A} {156, 123}

\bibitem[\protect\citeauthoryear{{Atek} et~al.,}{{Atek}
  et~al.}{2023}]{Atek2022}
{Atek} H.,  et~al., 2023, \mn@doi [\mnras] {10.1093/mnras/stac3144}, \href
  {https://ui.adsabs.harvard.edu/abs/2023MNRAS.519.1201A} {519, 1201}

\bibitem[\protect\citeauthoryear{{Barbary}}{{Barbary}}{2016}]{Barbary2016}
{Barbary} K.,  2016, \mn@doi [The Journal of Open Source Software]
  {10.21105/joss.00058}, \href
  {https://ui.adsabs.harvard.edu/abs/2016JOSS....1...58B} {1, 58}

\bibitem[\protect\citeauthoryear{Barden, Häußler, Peng, McIntosh  \&
  Guo}{Barden et~al.}{2012}]{barden_galapagos_2012}
Barden M.,  Häußler B.,  Peng C.~Y.,  McIntosh D.~H.,   Guo Y.,  2012,
  \mn@doi [\mnras] {10.1111/j.1365-2966.2012.20619.x}, 422, 449

\bibitem[\protect\citeauthoryear{{Bertin} \& {Arnouts}}{{Bertin} \&
  {Arnouts}}{1996}]{Bertin1996}
{Bertin} E.,  {Arnouts} S.,  1996, \mn@doi [\aaps] {10.1051/aas:1996164}, \href
  {https://ui.adsabs.harvard.edu/abs/1996A&AS..117..393B} {117, 393}

\bibitem[\protect\citeauthoryear{{Boyer} et~al.,}{{Boyer}
  et~al.}{2022}]{Boyer2022}
{Boyer} M.~L.,  et~al., 2022, \mn@doi [Research Notes of the American
  Astronomical Society] {10.3847/2515-5172/ac923a}, \href
  {https://ui.adsabs.harvard.edu/abs/2022RNAAS...6..191B} {6, 191}

\bibitem[\protect\citeauthoryear{{Bradley} et~al.,}{{Bradley}
  et~al.}{2022}]{larry_bradley_2022_6825092}
{Bradley} L.,  et~al., 2022, {astropy/photutils: 1.5.0}, Zenodo,
  \mn@doi{10.5281/zenodo.6825092}

\bibitem[\protect\citeauthoryear{{Brammer} \& {Matharu}}{{Brammer} \&
  {Matharu}}{2021}]{Brammer2021}
{Brammer} G.,  {Matharu} J.,  2021, {gbrammer/grizli: Release 2021}, Zenodo,
  \mn@doi{10.5281/zenodo.5012699}

\bibitem[\protect\citeauthoryear{{Brammer}, {van Dokkum}  \& {Coppi}}{{Brammer}
  et~al.}{2008}]{Brammer2008}
{Brammer} G.~B.,  {van Dokkum} P.~G.,   {Coppi} P.,  2008, \mn@doi [\apj]
  {10.1086/591786}, \href
  {https://ui.adsabs.harvard.edu/abs/2008ApJ...686.1503B} {686, 1503}

\bibitem[\protect\citeauthoryear{{Brinchmann}}{{Brinchmann}}{2022}]{Brinchmann2022}
{Brinchmann} J.,  2022, \mn@doi [arXiv e-prints] {10.48550/arXiv.2208.07467},
  \href {https://ui.adsabs.harvard.edu/abs/2022arXiv220807467B} {p.
  arXiv:2208.07467}

\bibitem[\protect\citeauthoryear{Bruzual \& Charlot}{Bruzual \&
  Charlot}{2003}]{BruzualCharlot2003}
Bruzual G.,  Charlot S.,  2003, \mn@doi [Monthly Notices of the Royal
  Astronomical Society] {10.1046/j.1365-8711.2003.06897.x}, 344, 1000

\bibitem[\protect\citeauthoryear{{Calzetti}, {Armus}, {Bohlin}, {Kinney},
  {Koornneef}  \& {Storchi-Bergmann}}{{Calzetti} et~al.}{2000}]{Calzetti2000}
{Calzetti} D.,  {Armus} L.,  {Bohlin} R.~C.,  {Kinney} A.~L.,  {Koornneef} J.,
   {Storchi-Bergmann} T.,  2000, \mn@doi [\apj] {10.1086/308692}, \href
  {https://ui.adsabs.harvard.edu/abs/2000ApJ...533..682C} {533, 682}

\bibitem[\protect\citeauthoryear{{Caminha}, {Suyu}, {Mercurio}, {Brammer},
  {Bergamini}, {Acebron}  \& {Vanzella}}{{Caminha} et~al.}{2022}]{Caminha2022}
{Caminha} G.~B.,  {Suyu} S.~H.,  {Mercurio} A.,  {Brammer} G.,  {Bergamini} P.,
   {Acebron} A.,   {Vanzella} E.,  2022, \mn@doi [\aap]
  {10.1051/0004-6361/202244517}, \href
  {https://ui.adsabs.harvard.edu/abs/2022A&A...666L...9C} {666, L9}

\bibitem[\protect\citeauthoryear{{Carnall} et~al.,}{{Carnall}
  et~al.}{2023}]{Carnall2022}
{Carnall} A.~C.,  et~al., 2023, \mn@doi [\mnras] {10.1093/mnrasl/slac136},
  \href {https://ui.adsabs.harvard.edu/abs/2023MNRAS.518L..45C} {518, L45}

\bibitem[\protect\citeauthoryear{{Coe} et~al.,}{{Coe} et~al.}{2019}]{Coe2019}
{Coe} D.,  et~al., 2019, \mn@doi [\apj] {10.3847/1538-4357/ab412b}, \href
  {https://ui.adsabs.harvard.edu/abs/2019ApJ...884...85C} {884, 85}

\bibitem[\protect\citeauthoryear{Coleman, Wu  \& Weedman}{Coleman
  et~al.}{1980}]{coleman_colors_1980}
Coleman G.~D.,  Wu C.-C.,   Weedman D.~W.,  1980, \mn@doi [\apjs]
  {10.1086/190674}, 43, 393

\bibitem[\protect\citeauthoryear{{Conroy} \& {Gunn}}{{Conroy} \&
  {Gunn}}{2010}]{Conroy2010}
{Conroy} C.,  {Gunn} J.~E.,  2010, \mn@doi [\apj]
  {10.1088/0004-637X/712/2/833}, \href
  {https://ui.adsabs.harvard.edu/abs/2010ApJ...712..833C} {712, 833}

\bibitem[\protect\citeauthoryear{{Conroy}, {Gunn}  \& {White}}{{Conroy}
  et~al.}{2009}]{Conroy2009}
{Conroy} C.,  {Gunn} J.~E.,   {White} M.,  2009, \mn@doi [\apj]
  {10.1088/0004-637X/699/1/486}, \href
  {https://ui.adsabs.harvard.edu/abs/2009ApJ...699..486C} {699, 486}

\bibitem[\protect\citeauthoryear{{Curti} et~al.,}{{Curti}
  et~al.}{2023}]{Curti2023}
{Curti} M.,  et~al., 2023, \mn@doi [\mnras] {10.1093/mnras/stac2737}, \href
  {https://ui.adsabs.harvard.edu/abs/2023MNRAS.518..425C} {518, 425}

\bibitem[\protect\citeauthoryear{Daddi, Cimatti, Renzini, Fontana, Mignoli,
  Pozzetti, Tozzi  \& Zamorani}{Daddi et~al.}{2004}]{daddi_new_2004}
Daddi E.,  Cimatti A.,  Renzini A.,  Fontana A.,  Mignoli M.,  Pozzetti L.,
  Tozzi P.,   Zamorani G.,  2004, \mn@doi [\apj] {10.1086/425569}, 617, 746

\bibitem[\protect\citeauthoryear{{Doyon} et~al.,}{{Doyon}
  et~al.}{2012}]{Doyon2012}
{Doyon} R.,  et~al., 2012, in {Clampin} M.~C.,  {Fazio} G.~G.,  {MacEwen}
  H.~A.,   {Oschmann} Jacobus~M. J.,  eds,  Society of Photo-Optical
  Instrumentation Engineers (SPIE) Conference Series Vol. 8442, Space
  Telescopes and Instrumentation 2012: Optical, Infrared, and Millimeter Wave.
  p. 84422R, \mn@doi{10.1117/12.926578}

\bibitem[\protect\citeauthoryear{{Draine} \& {Li}}{{Draine} \&
  {Li}}{2007}]{Draine2007}
{Draine} B.~T.,  {Li} A.,  2007, \mn@doi [\apj] {10.1086/511055}, \href
  {https://ui.adsabs.harvard.edu/abs/2007ApJ...657..810D} {657, 810}

\bibitem[\protect\citeauthoryear{Drlica-Wagner et~al.,}{Drlica-Wagner
  et~al.}{2018}]{Drlica-Wagner+2018}
Drlica-Wagner A.,  et~al., 2018, \mn@doi [The Astrophysical Journal Supplement
  Series] {10.3847/1538-4365/aab4f5}, 235, 33

\bibitem[\protect\citeauthoryear{{Euclid Collaboration: Desprez}
  et~al.,}{{Euclid Collaboration: Desprez} et~al.}{2020}]{Desprez2020}
{Euclid Collaboration: Desprez} G.,  et~al., 2020, \mn@doi [\aap]
  {10.1051/0004-6361/202039403}, \href
  {https://ui.adsabs.harvard.edu/abs/2020A&A...644A..31E} {644, A31}

\bibitem[\protect\citeauthoryear{{Ferland} et~al.,}{{Ferland}
  et~al.}{2017}]{Ferland2017}
{Ferland} G.~J.,  et~al., 2017, \rmxaa, \href
  {https://ui.adsabs.harvard.edu/abs/2017RMxAA..53..385F} {53, 385}

\bibitem[\protect\citeauthoryear{{Ferreira} et~al.,}{{Ferreira}
  et~al.}{2022}]{Ferreira2022}
{Ferreira} L.,  et~al., 2022, \mn@doi [\apjl] {10.3847/2041-8213/ac947c}, \href
  {https://ui.adsabs.harvard.edu/abs/2022ApJ...938L...2F} {938, L2}

\bibitem[\protect\citeauthoryear{Fitzpatrick}{Fitzpatrick}{1999}]{fitzpatrick_correcting_1999}
Fitzpatrick E.~L.,  1999, \mn@doi [\pasp] {10.1086/316293}, 111, 63

\bibitem[\protect\citeauthoryear{{Galametz} et~al.,}{{Galametz}
  et~al.}{2013}]{Galametz2013}
{Galametz} A.,  et~al., 2013, \mn@doi [\apjs] {10.1088/0067-0049/206/2/10},
  \href {https://ui.adsabs.harvard.edu/abs/2013ApJS..206...10G} {206, 10}

\bibitem[\protect\citeauthoryear{{Golubchik}, {Furtak}, {Meena}  \&
  {Zitrin}}{{Golubchik} et~al.}{2022}]{Golubchik2022}
{Golubchik} M.,  {Furtak} L.~J.,  {Meena} A.~K.,   {Zitrin} A.,  2022, \mn@doi
  [\apj] {10.3847/1538-4357/ac8ff1}, \href
  {https://ui.adsabs.harvard.edu/abs/2022ApJ...938...14G} {938, 14}

\bibitem[\protect\citeauthoryear{{Gray} \& {Corbally}}{{Gray} \&
  {Corbally}}{2009}]{GrayCorballyBook}
{Gray} R.~O.,  {Corbally} C.~J.,  2009, {Stellar Spectral Classification}.
Princeton University Press, Princeton, New Jersey, USA

\bibitem[\protect\citeauthoryear{Guo et~al.,}{Guo
  et~al.}{2013}]{guo_candels_2013}
Guo Y.,  et~al., 2013, \mn@doi [\apjs] {10.1088/0067-0049/207/2/24}, 207, 24

\bibitem[\protect\citeauthoryear{{Harris} et~al.,}{{Harris}
  et~al.}{2020}]{Harris2020}
{Harris} C.~R.,  et~al., 2020, \mn@doi [\nat] {10.1038/s41586-020-2649-2},
  \href {https://ui.adsabs.harvard.edu/abs/2020Natur.585..357H} {585, 357}

\bibitem[\protect\citeauthoryear{{Hunter}}{{Hunter}}{2007}]{Hunter2007}
{Hunter} J.~D.,  2007, \mn@doi [Computing in Science and Engineering]
  {10.1109/MCSE.2007.55}, \href
  {https://ui.adsabs.harvard.edu/abs/2007CSE.....9...90H} {9, 90}

\bibitem[\protect\citeauthoryear{{Ilbert} et~al.,}{{Ilbert}
  et~al.}{2009}]{Ilbert2009}
{Ilbert} O.,  et~al., 2009, \mn@doi [\apj] {10.1088/0004-637X/690/2/1236},
  \href {https://ui.adsabs.harvard.edu/abs/2009ApJ...690.1236I} {690, 1236}

\bibitem[\protect\citeauthoryear{{Ilbert} et~al.,}{{Ilbert}
  et~al.}{2013}]{Ilbert2013}
{Ilbert} O.,  et~al., 2013, \mn@doi [\aap] {10.1051/0004-6361/201321100}, \href
  {http://adsabs.harvard.edu/abs/2013A%26A...556A..55I} {556, A55}

\bibitem[\protect\citeauthoryear{{Kennicutt Jr.}}{{Kennicutt
  Jr.}}{1998}]{Kennicutt1998}
{Kennicutt Jr.} R.~C.,  1998, \mn@doi [\araa] {10.1146/annurev.astro.36.1.189},
  \href {https://ui.adsabs.harvard.edu/abs/1998ARA&A..36..189K} {36, 189}

\bibitem[\protect\citeauthoryear{{Kroupa}}{{Kroupa}}{2001}]{Kroupa2001}
{Kroupa} P.,  2001, \mn@doi [\mnras] {10.1046/j.1365-8711.2001.04022.x}, \href
  {https://ui.adsabs.harvard.edu/abs/2001MNRAS.322..231K} {322, 231}

\bibitem[\protect\citeauthoryear{{Li} et~al.,}{{Li} et~al.}{2022}]{Li2022}
{Li} M.,  et~al., 2022, arXiv e-prints, \href
  {https://ui.adsabs.harvard.edu/abs/2022arXiv221101382L} {p. arXiv:2211.01382}

\bibitem[\protect\citeauthoryear{{Madau}}{{Madau}}{1995}]{Madau1995}
{Madau} P.,  1995, \mn@doi [\apj] {10.1086/175332}, \href
  {https://ui.adsabs.harvard.edu/abs/1995ApJ...441...18M} {441, 18}

\bibitem[\protect\citeauthoryear{{Mahler} et~al.,}{{Mahler}
  et~al.}{2022}]{Mahler2022arXiv}
{Mahler} G.,  et~al., 2022, arXiv e-prints, \href
  {https://ui.adsabs.harvard.edu/abs/2022arXiv220707101M} {p. arXiv:2207.07101}

\bibitem[\protect\citeauthoryear{Matharu \& Brammer}{Matharu \&
  Brammer}{2022}]{grizliconf2022}
Matharu J.,  Brammer G.,  2022, {Updated Configuration files for JWST NIRISS
  Slitless Spectroscopy}, \mn@doi{10.5281/zenodo.7628094}, \url
  {https://doi.org/10.5281/zenodo.7628094}

\bibitem[\protect\citeauthoryear{{Mowla} et~al.,}{{Mowla}
  et~al.}{2022}]{Mowla2022}
{Mowla} L.,  et~al., 2022, \mn@doi [\apjl] {10.3847/2041-8213/ac90ca}, \href
  {https://ui.adsabs.harvard.edu/abs/2022ApJ...937L..35M} {937, L35}

\bibitem[\protect\citeauthoryear{{Polletta} et~al.,}{{Polletta}
  et~al.}{2007}]{Polletta2007}
{Polletta} M.,  et~al., 2007, \mn@doi [\apj] {10.1086/518113}, \href
  {https://ui.adsabs.harvard.edu/abs/2007ApJ...663...81P} {663, 81}

\bibitem[\protect\citeauthoryear{{Pontoppidan} et~al.,}{{Pontoppidan}
  et~al.}{2022}]{Pontoppidan2022}
{Pontoppidan} K.~M.,  et~al., 2022, \mn@doi [\apjl] {10.3847/2041-8213/ac8a4e},
  \href {https://ui.adsabs.harvard.edu/abs/2022ApJ...936L..14P} {936, L14}

\bibitem[\protect\citeauthoryear{{Prevot}, {Lequeux}, {Maurice}, {Prevot}  \&
  {Rocca-Volmerange}}{{Prevot} et~al.}{1984}]{Prevot1984}
{Prevot} M.~L.,  {Lequeux} J.,  {Maurice} E.,  {Prevot} L.,
  {Rocca-Volmerange} B.,  1984, \aap, \href
  {https://ui.adsabs.harvard.edu/abs/1984A&A...132..389P} {132, 389}

\bibitem[\protect\citeauthoryear{{Schaerer}, {Marques-Chaves}, {Barrufet},
  {Oesch}, {Izotov}, {Naidu}, {Guseva}  \& {Brammer}}{{Schaerer}
  et~al.}{2022}]{Schaerer2022}
{Schaerer} D.,  {Marques-Chaves} R.,  {Barrufet} L.,  {Oesch} P.,  {Izotov}
  Y.~I.,  {Naidu} R.,  {Guseva} N.~G.,   {Brammer} G.,  2022, \mn@doi [\aap]
  {10.1051/0004-6361/202244556}, \href
  {https://ui.adsabs.harvard.edu/abs/2022A&A...665L...4S} {665, L4}

\bibitem[\protect\citeauthoryear{Schlafly \& Finkbeiner}{Schlafly \&
  Finkbeiner}{2011}]{schlafly_measuring_2011}
Schlafly E.~F.,  Finkbeiner D.~P.,  2011, \mn@doi [\apj]
  {10.1088/0004-637X/737/2/103}, 737, 103

\bibitem[\protect\citeauthoryear{{Villaume}, {Conroy}  \& {Johnson}}{{Villaume}
  et~al.}{2015}]{Villaume2015}
{Villaume} A.,  {Conroy} C.,   {Johnson} B.~D.,  2015, \mn@doi [\apj]
  {10.1088/0004-637X/806/1/82}, \href
  {https://ui.adsabs.harvard.edu/abs/2015ApJ...806...82V} {806, 82}

\bibitem[\protect\citeauthoryear{{Virtanen} et~al.,}{{Virtanen}
  et~al.}{2020}]{Virtanen2020}
{Virtanen} P.,  et~al., 2020, \mn@doi [Nature Methods]
  {10.1038/s41592-019-0686-2}, \href
  {https://ui.adsabs.harvard.edu/abs/2020NatMe..17..261V} {17, 261}

\bibitem[\protect\citeauthoryear{{Willott} et~al.,}{{Willott}
  et~al.}{2022}]{willott2022}
{Willott} C.~J.,  et~al., 2022, \mn@doi [\pasp] {10.1088/1538-3873/ac5158},
  \href {https://ui.adsabs.harvard.edu/abs/2022PASP..134b5002W} {134, 025002}

\bibitem[\protect\citeauthoryear{{Yan}, {Ma}, {Ling}, {Cheng}  \&
  {Huang}}{{Yan} et~al.}{2023}]{Yan2022}
{Yan} H.,  {Ma} Z.,  {Ling} C.,  {Cheng} C.,   {Huang} J.-S.,  2023, \mn@doi
  [\apjl] {10.3847/2041-8213/aca80c}, \href
  {https://ui.adsabs.harvard.edu/abs/2023ApJ...942L...9Y} {942, L9}

\bibitem[\protect\citeauthoryear{York et~al.,}{York et~al.}{2000}]{York+2000}
York D.~G.,  et~al., 2000, \mn@doi [The Astronomical Journal] {10.1086/301513},
  120, 1579

\makeatother
\end{thebibliography}




\appendix

\section{{\it JWST} / NIRSpec redshifts}\label{sec:methods-nirspecredshifts}

We re-analyse the {\it JWST}/NIRSpec EROs of the SMACS~J0723.3$-$7327 field obtained from the two {\it JWST}/NIRSpec microshutter array pointings ({\tt s007} and {\tt s008}). Both pointings use two grating-filter combinations:
G235M/F170LP and G395M/F290LP, which provide a $1.75$--$5.20~\mu$m wavelength coverage with a spectral resolution of R$\sim$1000. We use the level-3 calibration data products retrieved from MAST which are processed with version 1.5.3 of the {\it JWST} Science Calibration Pipeline and the  {\tt jwst\_0916.pmap} calibration reference data.

After processing of the {\it JWST}/NIRSpec data, a redshift catalogue is created (see Table~\ref{tab:nirspec}). This catalogue contains 15 secure redshift measurements, two probable ones, three unsecured ones, and two stars. We follow the same flagging scheme as in \citet{Caminha2022}: we use a flag of 3 for a secure redshifts, 2 for probable, 9 for unsecured, and 4 for stars. We describe here the main steps of the processing.

The first step consists in a visual inspection (GD) of the 2D spectra to search for emission lines. Out of the 35 spectra available, 16 contain at least two emission lines, two only have one emission line, and no emission lines are found in the remaining 17.

For 15 of the spectra with multiple emission lines, we are able to securely identify the nature of the lines, leading to secure redshift estimates ($\rm flag=3$). These estimates are then compared and validated to the redshifts already provided in the literature \citep[e.g.,][]{Schaerer2022,Carnall2022}. However, we are not able to find consistent line identifications for the remaining spectrum with multiple features ({\tt s06113}). {The photometric redshift for this source is $z_{\rm phot}=0.44$ which does not match any line identified.} Therefore, we assume that the strongest emission line seen in the spectrum is H$\alpha$ and we flag the source with an unsecured redshift ($\rm flag=9$).

Spectra with single emission lines visible in the 2D spectra {follow the same treatment as source {\tt s06113} and} are also flagged with unsecured redshifts ($\rm flag=9$), and the redshifts reported in Table~\ref{tab:nirspec} assume that the lines seen in the data are H$\alpha$. However, inspection of the imaging data shows that one of these sources ({\tt s08498}) is a multiply imaged arc, identified as object 4.1
in \citet{Caminha2022} and \citet{Mahler2022arXiv}. Since the redshift estimate based on the single line being H$\alpha$ (i.e., $z=2.211$) agrees with the prediction of geometric redshift from the two lensing models, we use a flag of 2 (probable redshift) for this source (instead of 9). Also, we attribute the same redshift and flag to the {\tt s08883} source, as it is one of the less amplified counter-images (image 4.2 in the lensing models) of the same source, even though no line is visible in its {\it JWST}/NIRSpec spectrum. Source {\tt s08886} follows the same treatment, but we keep a flag of $9$ since the source position is located on a substructure at the edge of the counter-images. Finally, the inspection of the images shows that two of the sources with no lines ({\tt s08311} and {\tt s10444}) are stars. These two sources are thus flagged as such ($\rm flag=4$).

\begin{table*}
    \centering
    \caption{List of redshifts we extract from the {\it JWST}/NIRSpec ERO data.  Quality flags are:  3 -- secure redshift, 2 -- probable redshift, 9 -- insecure redshift, and 4 -- star. Only the probable and secure redshifts are used in this work.}
    \begin{tabular}{l c c c c l}
        \hline
         ID & R.A. & Dec. & z & quality flag & First reported in \\
         \hline
         s01917 & 110.8711215748463 & -73.46572380454734 & 1.244 & 3 & \citet{Carnall2022} \\
         s02038	& 110.8713165212675	& -73.46477523235114 & 5.086 &	9 & This work\\
         s03042	& 110.7533321732116 & -73.45770341069363 &  1.996 &	3 & This work; {see also \citet{Brinchmann2022}}\\
         s03772 & 110.746055285353\phantom{0} & -73.45327524433173 & 0.424	& 3 & This work\\
         s04580	& 110.7946173689543 & -73.44916191536561 & 5.173 &	3 & \citet{Mahler2022arXiv} \\
         s04590	& 110.8593287020751	& -73.44916559196156 & 8.498 &	3 &	\citet{Schaerer2022} \\
         s05144 & 110.8396739029237 & -73.44535697626878 & 6.383 &	3 & \citet{Carnall2022} \\
         s05735	& 110.81500918475\phantom{00} & -73.4401668800989\phantom{0} & 1.508 & 3 & This work; {see also \citet{Brinchmann2022}}\\
         s06113	& 110.845619147255\phantom{0}	& -73.43715272896384 & 3.714 &	9 & This work\\
         s06355	& 110.8445941696538	& -73.4350589621277\phantom{0} & 7.665	& 3 & \citet{Schaerer2022} \\
         s08140	& 110.7880022389405	& -73.46186797161636 & 5.275 & 3 & \citet{Carnall2022} \\
         s08311	& 110.8285704845804	& -73.46034461257591 & 0.0\phantom{00} & 4 & This work\\
         s08498	& 110.8065345478661	& -73.45825402810799 & 2.211 & 2 & This work\\
         s08506	& 110.9165442177316	& -73.45881278128644 & 2.213 &	3 & \citet{Carnall2022} \\
         s08883	& 110.8052128959135	& -73.45462052060702 & 2.211 &	2 & This work \\
         s08886	& 110.8046666778851	& -73.45447964066358 & 2.211 &	9 & This work\\
         s09239	& 110.765980906207\phantom{0}	& -73.45165829289732 & 2.463 & 3 & \citet{Carnall2022} \\
         s09483	& 110.7975811991304	& -73.44906450342938 & 1.163 & 3 & \citet{Carnall2022} \\
         s09721	& 110.7904591922968	& -73.4470750831021\phantom{0} & 2.118	& 3 & This work; {see also \citet{Brinchmann2022}}\\
         s09922	& 110.8597563507448	& -73.44421005964534 & 2.743 & 3 & \citet{Carnall2022} \\
         s10444	& 110.8139003331693	& -73.4379320696678\phantom{0} & 0.0\phantom{00} & 4 & This work\\
         s10612	& 110.8339649064393	& -73.43452316524234 & 7.663 & 3 & \citet{Schaerer2022} \\
         \hline
    \end{tabular}
    \label{tab:nirspec}
\end{table*}
\footnotesize{Also independently confirmed in \citet{Brinchmann2022} with consistent redshift}

\begin{figure*}
    \centering
    \includegraphics[width=\linewidth]{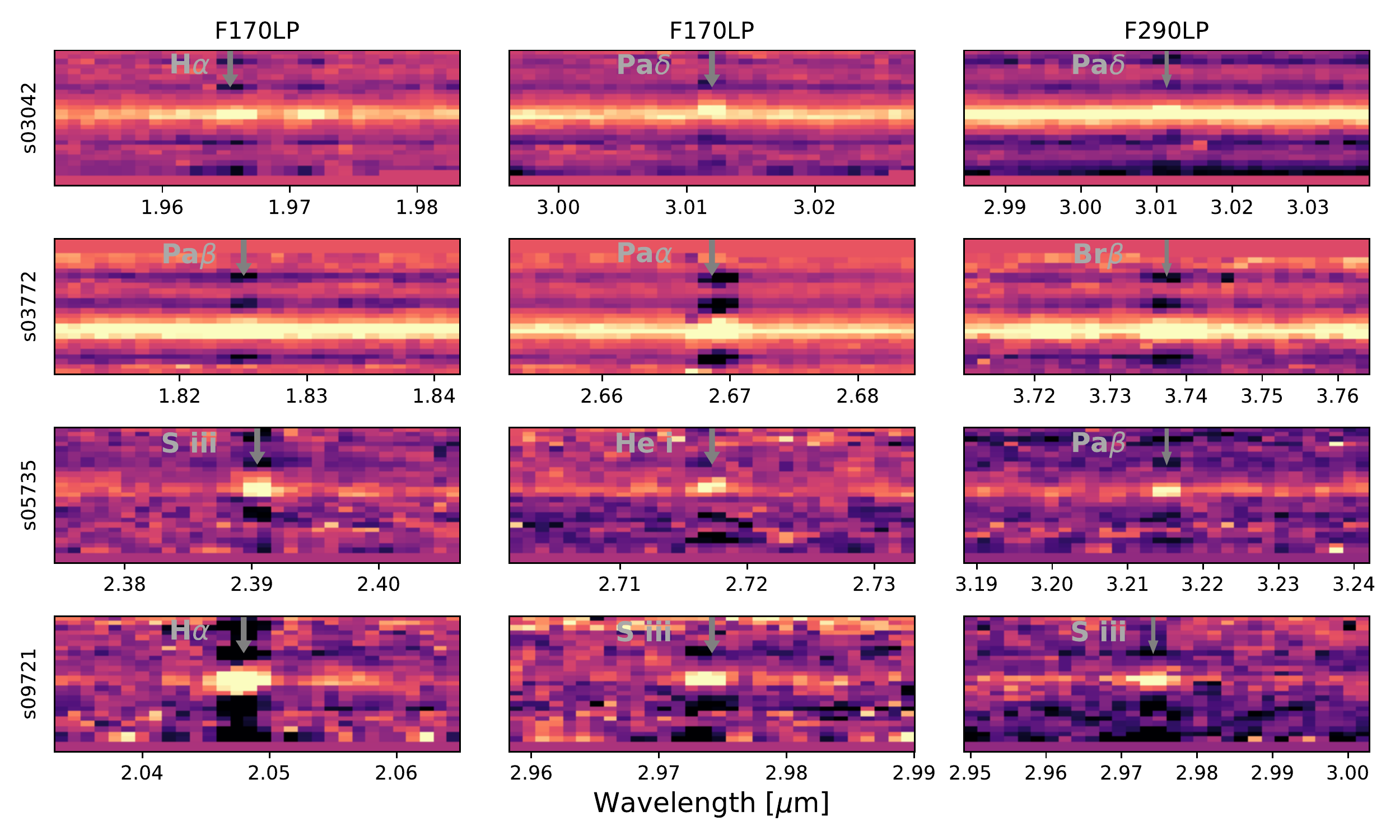}
    \caption{2D {\it JWST}/NIRSpec spectral lines detected and identified for the 4 new secure ($\rm flag = 3$) redshifts we obtain in this work.}
    \label{fig:nirspec}
\end{figure*}

\section{Description of our catalogues}
\label{sec:catalog_decription}

Table~\ref{tab:catalog} \&~\ref{tab:catalog2} describe the content of our final spectroscopic catalogue and of our multi-component catalogue, respectively.


\begin{table*}
	\centering
	\caption{Description of the main spectroscopic redshift catalogue.}
	\label{tab:catalog}
	\begin{tabular}{lcll} 
		\hline
		parameter &  units & catalog name & description \\
		\hline
		ID & & {\tt ID} & Object ID from the photometric catalog of Sec.~\ref{sec:photometry}\\
		RA & degrees & {\tt RA} & J2000 coordinates\\
		Decl & degrees & {\tt DEC} & J2000 coordinates \\
		$f_{\nu,F435W}$ & nJy & {\tt FLUX\_F435W} & Kron flux density in the F435W band \\
		        $\sigma_{\nu,F435W}$ & nJy & {\tt FLUXERR\_F435W} & Uncertainty in the F435W flux density \\
		$f_{\nu,F606W}$ & nJy & {\tt FLUX\_F606W}  & Kron flux density in the F606W band \\
				$\sigma_{\nu,F606W}$ & nJy & {\tt FLUXERR\_F606W}  &  Uncertainty in the F606W flux density\\
		$f_{\nu,F814W}$ & nJy & {\tt FLUX\_F814W}  & Kron flux density in the F814W band \\
				$\sigma_{\nu,F814W}$ & nJy & {\tt FLUXERR\_F814W}  &  Uncertainty in the F814W flux density\\
		$f_{\nu,F090W}$ & nJy & {\tt FLUX\_F090W}  & Kron flux density in the F090W band \\
				$\sigma_{\nu,F090W}$ & nJy & {\tt FLUXERR\_F090W}  &  Uncertainty in the F090W flux density\\
		$f_{\nu,F150W}$ & nJy & {\tt FLUX\_F150W}  & Kron flux density in the F150W band \\
				$\sigma_{\nu,F150W}$ & nJy & {\tt FLUXERR\_F150W}  & Uncertainty in the F150W flux density\\
		$f_{\nu,F200W}$ & nJy & {\tt FLUX\_F200W}  & Kron flux density in the F200W band \\
				$\sigma_{\nu,F200W}$ & nJy & {\tt FLUXERR\_F200W}  &  Uncertainty in the F200W flux density\\
		$f_{\nu,F277W}$ & nJy & {\tt FLUX\_F277W}  & Kron flux density in the F277W band \\
				$\sigma_{\nu,F277W}$ & nJy & {\tt FLUXERR\_F277W}  &  Uncertainty in the F277W flux density\\
		$f_{\nu,F356W}$ & nJy & {\tt FLUX\_F356W}  & Kron flux density in the F356W band \\
				$\sigma_{\nu,F356W}$ & nJy & {\tt FLUXERR\_F356W}  &  Uncertainty in the F356W flux density\\
		$f_{\nu,F444W}$ & nJy & {\tt FLUX\_F444W}  & Kron flux density in the F444W band \\
				$\sigma_{\nu,F444W}$ & nJy & {\tt FLUXERR\_F444W}  &  Uncertainty in the F444W flux density\\
		$z_{\rm grism}$ & & {\tt Z\_GRISM} & Grism redshift{; for multi-component sources, this is the mean of the different components}\\
		$\delta z_{\rm grism}$ & & {\tt ZERR\_GRISM} & {$1\sigma$ grism redshift uncertainty from {\tt Grizli}'s redshift fitting (random error), for multi-}\\
   & & & {component sources, this is the propagated error from the different components}\\
		Q-flag & & {\tt Q\_FLAG } & Grism quality flag: 1=Multiple features, 2=Single feature\\
		$z_{\rm spec}$ & & {\tt Z\_SPEC} & Spectroscopic redshift \\
		Spec. reference & & {\tt REF\_SPEC} & Origin of the $z_{\rm spec}$: `nirspec'={\it JWST}/NIRSpec; `muse'=VLT/MUSE  \\
		Grism merged flag & & {\tt FLAG\_MERGED\_GRISM} & Flag to identify merged grism sources\\
		Spec. merged flag & & {\tt FLAG\_MERGED\_SPEC} & Flag to identify merged spectroscopic sources\\
		$\mu$ & & {\tt MU} & Gravitational lensing magnification factor from the lens model of \citet{Caminha2022} \\
		\hline
	\end{tabular}
\end{table*}


\begin{table*}
    \centering
	\caption{Description of the multi-component source catalogue.}
	\label{tab:catalog2}
	\begin{tabular}{lcll}
	     \hline
		parameter &  units & catalogue name & description \\
		\hline
		ID & & {\tt ID} & Object ID; integer part denotes the source ID in the full catalogue, decimals denote the component IDs\\
		RA & degrees & {\tt RA} & J2000 coordinates\\
		Decl & degrees & {\tt DEC} & J2000 coordinates \\
		$z_{\rm grism}$ & & {\tt Z\_GRISM} & Grism redshift\\
		$\delta z_{\rm grism}$ & & {\tt ZERR\_GRISM} & {$1\sigma$ grism redshift uncertainty from {\tt Grizli}'s redshift fitting (random error)}\\
		Q-flag & & {\tt Q\_FLAG} & Grism quality flag: 1=Multiple features, 2=Single feature\\
		$z_{\rm spec}$ & & {\tt Z\_SPEC} & Spectroscopic redshift \\
		Spec. reference & & {\tt REF\_SPEC} & Origin of the $z_{\rm spec}$: `nirspec'={\it JWST}/NIRSpec; `muse'=VLT/MUSE  \\
		$\mu$ & & {\tt MU} & Gravitational lensing magnification factor from the lens model of \citet{Caminha2022} \\
		\hline
	\end{tabular}
\end{table*}

\section{Comparison with Li et al.}
\label{sec:Li-comparison}

\citet{Li2022} reports seven NIRISS grism redshifts in the SMACS~J0723.3$-$7327 field in the range $z_{\rm grism}=2.7-3.4$ as part of a larger analysis of the mass-metallicity relation of dwarf galaxies at cosmic noon. They determine their redshifts primarily based on the identification of the [OIII]$\lambda5007$ emission-line falling in the F200W filter at the redshifts reported in their work. Comparing with our independent analysis, we find the following:

\begin{itemize}
    
\item Three of the seven sources reported in \citet{Li2022} (IDs~1347, 1361, and 1424 in their sample) are also in our final catalogue (IDs~890, 894, and 919 in our catalogue, respectively), and the NIRISS grism redshifts measured in their work are consistent with ours. One of these three sources (ID~894) also has a NIRSpec redshift, and this redshift is in agreement with the NIRISS measurements.

\item One of the \citet{Li2022} sources (ID~50) is rejected in our analysis because it was observed with only one of the two orthogonal grisms, and the source is therefore deemed insecure in our work.

\item Two of the \citet{Li2022} sources (IDs~1475 and 1521) are rejected because they have no photometric redshifts that support the single feature detected in our analysis. Although our further inspection of the grism data suggests that there is some potential H$\beta$ detection in addition to the [OIII]$\lambda5007$ line, following our conservative methodology we do not include these sources in our catalogue.

\item One of the \citet{Li2022} sources (ID~1491) is rejected because our \EAZY and \Phosphoros redshift solutions disagree and suggest different identifications for the single line seen in the grism data (H$\alpha$ or [OIII]$\lambda5007$). Consequently, the nature of the line cannot be unambiguously determined per our methodology, and we therefore do not include this source in our conservative catalogue of secure spectroscopic redshifts.

\end{itemize}




\bsp	
\label{lastpage}
\end{document}